\documentclass[useAMS,usenatbib,a4paper]{mn2e}

\usepackage{graphicx}
\usepackage{wasysym}
\usepackage{hyperref}
\usepackage{subfig}
\usepackage{color}
\usepackage{verbatim}
\usepackage[Symbol]{upgreek}

\newcommand{\Mpc}{\,h^{-1}\,\mathrm{Mpc}}
\newcommand{\iMpc}{\,h\,\mathrm{Mpc}^{-1}}
\newcommand{\Msun}{\,h^{-1}\,\mathrm{M_\odot}}
\newcommand{\kpc}{\,h^{-1}\,\mathrm{kpc}}
\newcommand{\lcdm}{$\Lambda$CDM\ }
\newcommand{\tcdm}{$\tau$CDM\ }
\newcommand{\sinc}{\mathrm{sinc}\,}
\newcommand{\eg}{{e.g.}\ }

\newcommand{\etc}{{etc.}\ }
\newcommand{\sform}[2]{{#1}\times 10^{#2}}
\newcommand{\new}[1]{{\color{black}{#1}}}
\newcommand{\Om}{\Omega_\mathrm{m}}
\newcommand{\Ov}{\Omega_\Lambda}
\newcommand{\Ob}{\Omega_\mathrm{b}}
\renewcommand{\pi}{\uppi}

\def\gtsim{\mathrel{\lower0.6ex\hbox{$\buildrel {\textstyle >}
 \over {\scriptstyle \sim}$}}}
\def\ltsim{\mathrel{\lower0.6ex\hbox{$\buildrel {\textstyle <}
 \over {\scriptstyle \sim}$}}}

\def\m@th{\mathsurround=0pt }
\def\eqalign#1{\null\,\vcenter{\openup1\jot \m@th
 \ialign{\strut\hfil$\displaystyle{##}$&$\displaystyle{{}##}$\hfil
 \crcr#1\crcr}}\,}

\title[Remapping simulated halo catalogues in redshift space]{Remapping simulated halo catalogues in redshift space}
\author[A. J. Mead and J. A. Peacock]{A. J. Mead$^{1}$\thanks{E-mail: am@roe.ac.uk} and J. A. Peacock$^{1}$\\
$^{1}$Institute for Astronomy, University of Edinburgh, Royal Observatory, Blackford Hill, Edinburgh EH9 3HJ\\}

\begin{document}

\date{Accepted 2014 September 18.  Received 2014 September 18; in original form 2014 August 5}
\pagerange{\pageref{firstpage}--\pageref{lastpage}} 
\pubyear{2014}

\maketitle

\label{firstpage}

\begin{abstract}
We discuss the extension to redshift space of a rescaling algorithm, designed to alter the effective cosmology of a pre-existing simulated particle distribution or catalogue of dark matter haloes. The rescaling approach was initially developed by Angulo \& White and was adapted and applied to halo catalogues in real space in our previous work. This algorithm requires no information other than the initial and target cosmological parameters, and it contains no tuned parameters. It is shown here that the rescaling method also works well in redshift space, and that the rescaled simulations can reproduce the growth rate of cosmological density fluctuations appropriate for the target cosmology. Even when rescaling a grossly non-standard model with $\Lambda=0$ and zero baryons, the redshift-space power spectrum of standard $\Lambda$CDM can be reproduced to about 5\% error for $k<0.2\iMpc$. The ratio of quadrupole-to-monopole power spectra remains correct to the same tolerance up to $k=1\iMpc$, provided that the input halo catalogue contains measured internal velocity dispersions.
\end{abstract}


\section{Introduction}

Numerical simulations are now an essential standard tool in the analysis and interpretation of cosmological surveys. Modern examples of which deliver high statistical power and are seeking to measure cosmological parameters with ever-increasing precision, thus requiring a corresponding rise in the care with which observational selection effects are treated. Practical survey complications are inevitably encountered to some extent, such as difficulty in obtaining spectra for adjacent objects in multiplexed spectroscopic surveys, and these are most robustly treated via Monte Carlo: analysis of a simulated mock dataset where the underlying cosmological parameters are known. This approach is increasingly deployed in order to verify the robustness of parameter estimates, or to identify and eliminate small residual biases. It can be witnessed in action in measurements of Baryon Acoustic Oscillations (BAO) in the galaxy distribution (\eg \citealt{Anderson2014}) or in redshift-space distortions of galaxy clustering (\eg \citealt{Samushia2013}; \citealt{delaTorre2013a}).

\new{Such studies also require the generation of large libraries of simulations. An ensemble of mock datasets is always required for a given cosmology, simply in order to generate an estimate of the data covariance matrix; this is used on the assumption of a Gaussian error distribution to calculate the likelihood of a given hypothetical model with respect to the data. In order for the final parameter constraints to be meaningful (small `errors on the errors'), the data covariance matrix itself must be precisely measured. This can require an ensemble of $\sim1000$ simulations -- or significantly more if the size of the dataset under study is sufficiently large (\citealt{Taylor2013}).}

But the size of the numerical challenge is greatly increased if we accept that, in principle, a new ensemble of simulations is required for each new cosmological model under study.  Moreover, mock {\it galaxy\/} data are required, so in principle a galaxy formation code must be run as part of each simulation. The latter issue can be dealt with rapidly using halo occupation distribution (HOD) models (\eg \citealt{Zheng2005}), reducing the problem to one of obtaining catalogues of dark-matter haloes for the model under study.  A way to speed up this basic generation of dark-matter simulations was presented by \citeauthor{Angulo2010} (\citeyear{Angulo2010}; hereafter AW10). They showed that it was possible to rescale an $N$-body particle distribution so that the results closely approximated the output of a simulation with a different set of cosmological parameters.

In \citeauthor{Mead2014} (\citeyear{Mead2014}; hereafter MP14) we showed that the AW10 method could be applied directly to halo catalogues, yielding a mass function and power spectrum of haloes that were well matched to the target cosmology following rescaling. In fact, this approach is capable of yielding more accurate results, since it retains the mass-dependent halo bias relation and can incorporate a cosmology dependence of halo concentrations. Moreover, this method has a major practical advantage in terms of data compression: halo catalogues take up several orders of magnitude less storage than the raw particle data, and thus it is often the case that only the halo catalogue is made public (or even stored) by major simulation projects such as that of \cite{Rasera2010}.

In MP14 we showed that these methods worked well on haloes in real space; in this paper we extend our approach to redshift space. The algorithm presented here consists of the following steps: (a) Initially the length and time units in the original halo catalogue are rescaled, in order to match the halo mass function, exactly as in the original AW10 algorithm. (b) In MP14 we showed that the particle or halo distribution itself may be used to compute the linear displacement field, from which we modify particle or halo positions and velocities so that they reproduce the correct large-scale clustering in the target cosmology. \cite{Eisenstein2007} showed how to recreate the displacement fields via the over-density field, and \cite{Padmanabhan2012} used a variant of this approach to improve the sharpness of the BAO feature in BOSS data. (c) Finally, we modify the halo internal physical and velocity structure directly -- by `reconstituting' the density profiles around haloes so that they have the correct sizes and internal structure for the target cosmology. In this way, the rescaled velocity field contains both the linear velocities that contribute coherent redshift-space distortions (\citealt{Kaiser1987}) and the post-linear effects that cause the `fingers-of-God' (FOG) distortion.

Our paper is set out as follows: In Section \ref{sec:method} we briefly review the AW10 and MP14 algorithms and explain the extensions required to get good results in redshift space. We discuss the cosmology dependence of the internal physical and velocity structure of haloes. In Section \ref{sec:simulations} we describe our simulations and our methods for generating halo catalogues. In Section \ref{sec:power} we discuss our conventions and method for generating power spectra in redshift space. In Section \ref{sec:results} we show results for the full redshift-space power spectrum as well as the monopole and quadrupole moments of this. Results are presented for matter, haloes, and particles within haloes, and we show that an accurate recovery of the growth rate may be made from the rescaled distributions. Finally we sum up in Section \ref{sec:summary}.

\section{Method}
\label{sec:method}

\subsection{Overview of rescaling}

We begin with a summary of the main features of the AW10 method and
the MP14 extension.  Quantities in the target cosmology are denoted
with primes while the original quantities are unprimed. The original
simulation output at redshift $z$, in a box of size $L$, is rescaled
to a target simulation at redshift $z'$, in a box of size $L'=sL$. For
a given $z'$, $s$ and $z$ are chosen so as to minimize the difference
in $\sigma^2(R)$ (the {\it linear-theory\/} variance in density averaged over
spheres of comoving radius $R$) between the two cosmologies.  The following cost function
is convenient:
\begin{equation}
\delta_{\mathrm{rms}}^2(s,z\mid z')=\frac{1}{\ln(R'_2/R'_1)}\int_{R'_1}^{R'_2}
\frac{\mathrm{d}R}{R}\left[1-\frac{\sigma(R/s,z)}{\sigma'(R,z')}\right]^2\ ,
\label{eq:minimise}
\end{equation}
with $R'=sR$.
$R_1$ and $R_2$ are chosen so as to relate to the physical scale of the least 
and most massive haloes in the original simulation via
\begin{equation}
M=\frac{4}{3}\pi R^3 \bar{\rho}\ ,
\label{eq:mass_to_R}
\end{equation}
where $\bar\rho$ is the mean comoving density. This approach is taken because the halo mass
function is approximately universal when expressed in terms of the
variable $\nu=\delta_\mathrm{c}/\sigma(R)$ where
$\delta_\mathrm{c}\simeq 1.686$ (\eg \citealt{Press1974};
\citealt{Sheth1999}; \citealt{Sheth2001}), and because the statistics
of the nonlinear density field are driven by the mass function of
haloes. In detail, the mass function displays deviations from
universality at up to around the 10\% level, so more accurate
results might be obtained by 
directly matching theoretical predictions for a non-universal 
mass function (\eg \citealt{Lukic2007}; \citealt{Reed2007};
\citealt{Tinker2008}). We have chosen not to do
this because these mass functions are tuned to specific cosmological
parameters and we are interested in quite broad shifts in cosmology in
this work. Similarly, one might minimize the difference in
$\nu(M)=\delta_\mathrm{c}/\sigma(M)$ where $\delta_\mathrm{c}$ could be
taken to vary with cosmology. For standard models, these
variations in $\delta_\mathrm{c}$ are negligible (\eg
\citealt{Eke1996}; \citealt{Lacey1993}; \citealt{Percival2005}), but
large variations in $\delta_\mathrm{c}$ are a feature of some modified
gravity models (\eg \new{\citealt{Schmidt2009a}; \citealt{Li2012b}}),
owing to screening mechanisms.

In order to conserve mass the scaling in length units simultaneously implies a scaling in mass:
\begin{equation}
M'= s^3\frac{\Om'}{\Om} M\equiv s_\mathrm{m} M\ .
\label{eq:mass_scaling}
\end{equation}
Note that we use units of $\Mpc$ for length and $\Msun$ for mass and the necessary factors of $h$ are included in the scalings. 

Additionally the dimensionless velocity units of the simulation must be conserved before and after scaling (see AW10; MP14) which implies a scaling in bulk velocities of particles or haloes such that
\begin{equation}
\mathbf{v}'=s\frac{H'(a')f'_\mathrm{g}a'}{H(a)f_\mathrm{g}a}\mathbf{v}\ ,
\label{eq:velocity_scaling}
\end{equation}
\new{where $H$ is the Hubble parameter at the epoch in question; $f_\mathrm{g}\equiv \mathrm{d}\ln g/\mathrm{d}\ln a$ is the logarithmic growth rate; $g(a)$ is the linear theory growth function; and $a$ is the scale factor.}

Following scaling in $s$ and $z$, the two cosmologies should have
similar nonlinear power spectra; but the linear power on large scales
will in general not be matched.
This difference can be corrected for by using
the approximation of \citeauthor{Zeldovich1970}
(\citeyear{Zeldovich1970}; hereafter ZA) to perturb the particle or
halo positions using the displacement field: the phase of each mode of
the large-scale displacement field is preserved, but the amplitude is altered to match the
target power spectrum.
The displacement field $\mathbf{f}$ is defined so as to move particles from their initial Lagrangian positions $\mathbf{q}$ to their comoving Eulerian positions $\mathbf{x}$: 
\begin{equation} 
\mathbf{x}=\mathbf{q}+\mathbf{f}\ .
\label{eq:q_to_x}
\end{equation} 
At linear order the displacement field is related to the matter over-density $\delta$ via 
\begin{equation} 
\delta=-\nabla\cdot\mathbf{f}\ , 
\end{equation} 
which in Fourier space is
\begin{equation} \mathbf{f}_{\mathbf{k}}=-i\frac{\delta_{\mathbf{k}}}{k^2}\mathbf{k}\ .
\label{eq:displacement}
\end{equation}
If the displacement field in the original simulation is known, then an additional displacement can be specified in Fourier space to reflect the differences in the linear matter power spectra between the two cosmologies:
\begin{equation}
\delta\mathbf{f}'_{\mathbf{k'}}=\left[\sqrt{\frac{\Delta_{\rm{lin}}^{'2}(k',z')}{\Delta_{\rm{lin}}^2(sk',z)}}-1\right]\mathbf{f}'_{\mathbf{k'}}\ .
\label{eq:move}
\end{equation}
where $\Delta^2$ is the power spectrum in the form of fractional variance in density
per $\ln k$, and
$\mathbf{f}'$ is the linear displacement 
field after the input simulation has been scaled.

The particle or halo positions in the original simulation can be used to estimate the overdensity field via equation (\ref{eq:displacement}). To derive the overdensity of matter from the halo overdensity field, $\delta_\mathrm{H}$, must be debiased, respecting the relation
\begin{equation} 
\delta_\mathrm{H}=b(M)\delta\ ;
\label{eq:bias}
\end{equation}
in practice, we use the bias relations of \cite{Sheth1999}. We then define a
number-weighted effective bias for all haloes:
\begin{equation}
b_\mathrm{eff}=\frac{\int_{\nu_\mathrm{min}}^{\nu_\mathrm{max}} \mathrm{d}\nu\,
  b(\nu)f(\nu)/m(\nu)}{\int_{\nu_\mathrm{min}}^{\nu_\mathrm{max}} \mathrm{d}\nu\,
  f(\nu)/m(\nu)}\ ,
\label{eq:eff_bias}
\end{equation}
where $f(\nu)\,d\nu$ is the fraction of the total density contributed
by haloes in the range $d\nu$.  
For very low-mass haloes
$b(M)<1$ and the method could in principle yield an unphysical
negative mass density. In practice, this method is of interest for
large-volume simulations where haloes of extremely low mass are
not resolved, and we always find $b_\mathrm{eff}>1$.
Negative densities can always be avoided either by using something more
sophisticated than a linear biasing relation, or
by using haloes only above a certain
mass to recreate the overdensity field.

Equation (\ref{eq:displacement}) is only valid for the linear
components of both fields, so in practice the reconstructed $\delta$
must be smoothed with a window of width the nonlinear scale
$R_\mathrm{nl}$ to remove the nonlinear components. To do this we use
a Gaussian filter
\begin{equation}
F(k)=\mathrm{e}^{-k^2 R_\mathrm{nl}^2 /2}\ ,
\end{equation}
where $\sigma(R_\mathrm{nl},z)=1$.  Then equation
(\ref{eq:displacement}) can be used to estimate the linear
displacement field in Fourier space and particles can be
displaced differentially in order to account for the differing linear power
spectra:
\begin{equation}
\mathbf{x''}=\mathbf{x'}+\delta\mathbf{f}'\ ,
\label{eq:differential_displacements}
\end{equation}
where the double dash indicates positions after this displacement
has been applied. As noted in MP14, haloes are
biased tracers of the density field and their 
effective displacement fields must therefore
also be biased. Thus $\mathbf{f}_\mathrm{H}=b(M)\mathbf{f}$ for
each halo. In MP14, good results for the rescaled halo power spectrum
were in general \emph{not} obtained unless a biased displacement field was used.

The ZA also allows residual differences in linear velocities to be
corrected on a mode-by-mode basis. In the ZA the peculiar velocity
field ($\mathbf{v}\equiv a\dot{\mathbf{x}}$) is related to the
  displacement field by $\mathbf{v}=aHf_\mathrm{g}\mathbf{f}$ and
  additional differential changes to the peculiar velocities of
  particles or haloes are then given by
\begin{equation}
\delta\mathbf{v}'_\mathbf{k'}=a'H'f_\mathrm{g}'\left[\sqrt{\frac{\Delta_\mathrm{lin}^{'2}(k',z')}{\Delta_\mathrm{lin}^2(sk',z)}}-1\right]\mathbf{f}'_\mathbf{k'}\ .
\label{eq:move_v}
\end{equation}
Note that, in contrast to the displacement field case, the halo
velocities are unbiased: the equivalence principle requires haloes of
all masses in a given region to share a common large-scale velocity. The final
velocities after the displacement field step are then
\begin{equation}
\mathbf{v}''=\mathbf{v}'+\delta\mathbf{v}'\ .
\end{equation}

At this stage the linear power and mass function should be very close to the desired target. According to the halo model (\eg \citealt{Peacock2000}; \citealt{Seljak2000}; \citealt{Cooray2002}),  matching the mass function should also yield the correct quasi-linear clustering. Note that the rescaling method naturally includes effects such as halo exclusion (\citealt*{Smith2011a}) because it is based on rescaling an exact $N$-body calculation. However, the highly nonlinear part of the power spectrum is influenced by the internal structure of haloes, and this will also change if the cosmology is altered (\new{\eg \citealt{Navarro1997}; \citealt{Bullock2001}; \citealt{Dooley2014}}). Such effects will alter the mass distribution defined by the haloes, and will also be important in producing mock galaxy catalogues using HOD methods. Here, a given halo is typically assigned a central galaxy and a number of satellite galaxies that are taken to trace the density profile. It is therefore necessary to consider how the internal structure of haloes should be rescaled. Note that such an adjustment is not part of the original AW10 method, which is a further advantage of the current approach.

\new{AW10 was first applied directly to galaxy catalogues by \cite{Ruiz2011}. Differences in galaxy formation between cosmological models have been studied using this method by \cite{Guo2013} and the cosmological constraints one can derive from differences in galaxy clustering are discussed by \cite{Simha2013} (who used sub-halo abundance matching to populate rescaled catalogues with galaxies). In all cases only small scales were investigated and the large-scale ZA correction was not applied to the halo distribution. MP14 was the first work to successfully apply the ZA correction when working directly with a halo catalogue.}

\subsection{Rescaling halo properties}

\subsubsection{Halo catalogues}

Beyond positions, velocities and masses of haloes, there are a number of other properties that may plausibly be stored in a halo catalogue (see Table \ref{tab:catalogue_scalings}). Halo radial density profiles have been shown to be accurately approximated on average by the `NFW' profile of \cite*{Navarro1997}:
\begin{equation}
\rho(r)=\frac{\rho_\mathrm{N}}{(r/r_\mathrm{s})(1+r/r_\mathrm{s})^2}\ ,
\label{eq:nfw}
\end{equation}
where $r_\mathrm{s}$ is a scale radius and $\rho_\mathrm{N}$ is a normalization to obtain the correct halo mass. Although subsequent work (\citealt{Moore1999}; \citealt{Merritt2005}; \citealt{Merritt2006}) showed this form to be imperfect at small $r$, it will suffice for our present purpose: mock galaxies are either central at $r=0$ exactly, or satellites that tend to be found around $r_\mathrm{s}$, where the NFW approximation is good. The density profile is truncated at the virial radius $r_\mathrm{v}$, which will depend on the exact definition of a halo; this can be measured directly or inferred from the halo mass:
\begin{equation}
r_\mathrm{v}=\left(\frac{3M}{4\pi\Delta_\mathrm{v}\bar{\rho}}\right)^{1/3}\ ,
\label{eq:virial_radius}
\end{equation}
where $\Delta_\mathrm{v}$ is the average density of a halo with respect to the background matter density. We adopt the common value of $178$, justified by the near-universality of the halo mass function when defining haloes via the friends-of-friends algorithm with a cosmology-independent linking length.

\begin{table*}
\centering
\begin{tabular}{c c c c}
\hline\hline 
quantity & symbol & scaling & comments \\ [0.5ex] 
\hline
Positions & $\mathbf{x}$ & $s$ & Additionally use equation (\ref{eq:move}) for linear displacements \\
Velocities & $\mathbf{v}$ & $s\,H'f'_\mathrm{g}a'/Hf_\mathrm{g} a$ & Additionally use equation (\ref{eq:move_v}) for linear velocities \\
Particle or halo mass & $M$ & $s^3\,\Om'/\Om$ & -- \\
Virial radius & $r_\mathrm{v}$ & $s$ & Although this depends on how $c$ is defined \\
Halo concentration & $c$ & $c'_\mathrm{th}/c_\mathrm{th}$ & $c_\mathrm{th}$ (theoretical) computed from any $c(M)$ relation \\
Halo velocity dispersion & $\sigma_\mathrm{v}$ & $s\sqrt{\Om'/\Om}$ & Alternatively use equation (\ref{eq:sigma_v_better}) with $c$ dependence \\
Inertia tensor eigenvalues & $\lambda^2$ & $s^2$ & -- \\
Normalized inertia tensor eigenvectors & $\mathbf{w}$ & -- & No scaling because they are normalized \\
\hline
\end{tabular}
\caption{Scalings for various quantities that may be contained in a halo catalogue.}
\label{tab:catalogue_scalings}
\end{table*}

The density profile is therefore fully specified via a value for $r_\mathrm{s}$ or alternatively for the halo concentration $c=r_\mathrm{v}/r_\mathrm{s}$. There are many suggested relations for determining the concentration as a function of cosmological parameters (\eg \citealt{Navarro1997}; \citealt{Eke2001}; \citealt{Bullock2001}; \citealt{Neto2007}). We adopt the relations of \cite{Bullock2001} because they are couched in term of physical quantities such as halo formation time, rather than being empirical fitting functions that apply only to a limited range of models.

Apart from virial radii and concentration, additional quantities that may be stored
with a halo catalogue are the halo velocity dispersion, $\sigma_v$, and the
eigenvalues, $\lambda^2$, and eigenvectors, $\mathbf{w}$, of
the moment of inertia tensor, which gives a measure of halo
asphericity:
\begin{equation}
I_{ij}=\sum_{k=1}^{N}(x_{k,i}-\bar{x}_{i})(x_{k,j}-\bar{x}_{j})\ .
\end{equation}
Here $\mathbf{x}_k$ is position of the $k^\mathrm{th}$ halo particle 
with halo centre-of-mass (CM) position $\bar{\mathbf{x}}$, $k\in\{1,...,N\}$ and 
there are $N$ particles in each halo and $i,j\in\{1,2,3\}$ and label coordinates.
Diagonalizing this tensor provides the axial ratios of the halo
(via the eigenvalues) and the orientation of the halo (via the
eigenvectors). In practice, we use a minimum size of $N=100$ particles
for this estimation.

We now discuss how each of these may be scaled: Measured
concentrations may be scaled by a ratio of theoretical predictions
taken from a relation such as \cite{Bullock2001}; this corrects
the mean relation, and implicitly assumes that the fractional scatter
in concentrations of individual haloes has no strong cosmology dependence.
Virial radii will be scaled by a factor of $s$
because the virial overdensity criterion is independent
of cosmology. For the NFW profile, and assuming isotropic orbits,
the halo velocity dispersion is
\begin{equation}
\sigma_v^2 = \frac{GM}{3r_\mathrm{v}}\; \frac{c[1-1/(1+c)^2-2\ln(1+c)/(1+c)]}{2[\ln(1+c) - c/(1+c)]^2}\ ,
\label{eq:sigma_v_better}
\end{equation} 
which would give the scaling a mild concentration dependence. We chose not to use this more complicated formula for this initial investigation, adopting the simple approximation of 
\begin{equation}
\sigma^2_v\simeq \frac{GM}{3 r_\mathrm{v}}\ ,
\label{eq:sigma_v_basic}
\end{equation}
from which a scaling factor for $\sigma_v$ of $s\sqrt{\Om'/\Om}$ follows. \new{The difference in $\sigma_v$ between equations (\ref{eq:sigma_v_better}) and (\ref{eq:sigma_v_basic}) is only $\simeq 10\%$ for concentrations of interest in this work ($c<10$).} We do not attempt to solve the Jeans equation for the velocity dispersion as a function of radius (which would be possible given assumptions about orbital anisotropy). The limitations of halo scaling in redshift space that we uncover below would not be removed by such complications. The scalings we use in practice are summarized in Table \ref{tab:catalogue_scalings}.

\subsubsection{Reconstitution of haloes}

Given a hypothesis for the the density profile of a given halo, one
may undertake {\it reconstitution\/} of the halo, either to
recreate the dark-matter particles of which the halo is composed, or
to populate it with mock galaxies. In the former case, one can use the
NFW profile to place particles at random in order to sample the density
(ignoring subhaloes).
The boundary of a halo can either be taken to be given
by $r_\mathrm{v}$ according to equation (\ref{eq:virial_radius}) with
a concentration given by a relation such as \cite{Bullock2001} or
these can be taken from rescaled stored values. Haloes can then be assigned a
velocity dispersion via either equation (\ref{eq:sigma_v_better}) or
(\ref{eq:sigma_v_basic}). The simplest approach is then to take the
velocity components to be Gaussian distributed (see
\citealt{Kazantzidis2004}; \citealt{Wojtak2005} for limitations); we
examined the approximation of Gaussian internal halo velocities in testing and
found it to be accurate for our particular definition of haloes in our
simulations (see section \ref{sec:simulations}). 

If desired,
aspherical reconstitution can be achieved by distorting the halo particle
distribution once it has  been generated by the spherical halo reconstitution process
described above. If the square roots of the eigenvalues are
$\lambda_1$, $\lambda_2$ and $\lambda_3$ then each coordinate of the
particles in the reconstituted halo in the CM frame,
$\mathbf{y}$, is modified according to
\begin{equation}
y_i \rightarrow 3\lambda_i y_i/(\lambda_1+\lambda_2+\lambda_3)\ ,
\label{eq:asphericity}
\end{equation}
where $i\in\{1,2,3\}$. We also considered the prescription $y_i
\rightarrow \lambda_i y_i/(\lambda_1 \lambda_2 \lambda_3)^{1/3}$ \etc
but found this to work less well in recovering the shapes of
aspherical haloes. The relative to CM position vector of each halo particle is
then rotated by the inverse matrix of eigenvectors in order to orient
the halo correctly.

\subsubsection{Restructuring halo particles}
\label{sec:restructuring}

Given access to the full particle distribution, two options are
available to match the haloes in the target cosmology; one could
either remove haloes from the rescaled particle distribution entirely,
replacing them with reconstituted haloes in the manner described above
-- a method we call `regurgitation'. Alternatively the haloes can be
identified in the scaled distribution and reshaped to account for the
different cosmology -- a method we call `restructuring'. Again we
point out that it would be difficult to implement the biased
displacement field for haloes if working from a particle distribution,
and so the $b(M)$ relation will not be correctly reproduced in this
case.

To restructure haloes in a rescaled particle distribution one can proceed as follows: 
The amount of mass enclosed by an NFW profile at a radius $r$ is given by
\begin{equation}
M_\mathrm{enc}(r)=M\frac{\mu(r/r_\mathrm{s})}{\mu(c)}\equiv f(r) M\ ,
\label{eq:mass_enclosed}
\end{equation}
where
\begin{equation}
\mu(x)=\ln{(1+x)} - \frac{x}{1+x}\ .
\label{eq:nfw_f_factor}
\end{equation}
Haloes can be reshaped by the ratio of mass enclosed at a radius $r$ from the halo centre in each case. A scaled particle originally at $r'$ should be moved to $r''$, given by
\begin{equation}
r''=f''^{-1}[f'(r')]\ ,
\label{eq:inverse_f}
\end{equation}
where $f^{-1}$ indicates the inverse function. $f''$ will be the value of $f$ in the target cosmology whereas $f'$ will be the value in the original cosmology \emph{after it has been scaled}. Particle positions relative to the CM can then be reassigned via
\begin{equation}
\mathbf{y''}=\frac{r''}{r'}\mathbf{y'}\ ,
\label{eq:cm_pos_scaling}
\end{equation}
so that they end up with the correct interior distribution for haloes in the
new cosmology, but allowing the haloes to maintain any asphericity they may have had. This also
means that the haloes retain any dispersion in structure and environmental dependence 
that they may have had in the original
simulation. This requires the input of a $c(M)$ relation, and as before we adopt the
relations of \cite{Bullock2001}.
For `real' haloes it is possible for particles to be
outside the official virial radius (calculated from equation
\ref{eq:virial_radius}) for haloes defined with a friends-of-friends
(FOF) algorithm, which is used in this work. Using equation
(\ref{eq:cm_pos_scaling}) is still possible in this case but $f>1$ for
these particles.

Halo particle velocities, $\mathbf{u}$, relative to the CM velocity, can be reassigned via
\begin{equation}
\mathbf{u}''=\frac{\sigma''_v}{\sigma'_v}\mathbf{u}'\ ,
\label{eq:cm_vel_scaling}
\end{equation}
with $\sigma_v$ taken from either equation (\ref{eq:sigma_v_basic}) or (\ref{eq:sigma_v_better}). This should make the halo velocity dispersion appropriate for the new cosmology whilst maintaining the dispersion and environmental dependence in the dispersion relation for haloes of a given mass.

\section{Simulations}
\label{sec:simulations}

We now show the operation of halo rescaling in redshift space, using
the same matched set of
simulations, with corresponding halo catalogues, as in MP14.
The simulation parameters are given in Table \ref{tab:simulations}. The target
\lcdm simulation is a WMAP1-style cosmology (\citealt{Spergel2003})
run with the same transfer function as that of the Millennium
Simulation (\citealt{Springel2005}) which was generated using
\texttt{CMBFAST} (\citealt{Seljak1996}). The input simulation
\tcdm is a flat matter-only simulation run with a DEFW transfer
function (\citealt{Davis1985}) tuned to have a similar spectral shape
to that of the Millennium Simulation. The historical popularity
of \tcdm models is explained in MP14.

We take this extreme approach (no $\Lambda$ or baryons) in order
to see how well rescaling can work over any range of models
of conceivable interest. For smaller variations in cosmological
parameters, we may expect that the results will be correspondingly
more accurate. Additionally; any practical application of this method would
scale to a variety of cosmological models from a \emph{single} parent 
simulation of high $\sigma_8$, such that it explored a large range of fluctuation amplitudes.
In this case scalings to standard models are likely to come from the $\Omega_\mathrm{m}\simeq 1$
regime of the parent simulation anyway, even if it is $\Lambda$CDM.

Initial conditions were generated at $z_i=199$, using a glass
initial load of
$512^3$ particles, created using the \texttt{N-GenIC} code. The simulations
were run using the cosmological $N$-body code
\texttt{Gadget-2} of \cite{Gadget2}. As in MP14, we used
the same phases for the Fourier modes in
the target and original simulations, so that the approximate and exact
target halo fields can be compared visually, and not purely at the level
of power spectra. This also allows us to analyse the results of the
rescaling without the added complication of cosmic variance.

\begin{table*}
\centering
\begin{tabular}{c c c c c c c c c c}
\hline\hline 
Simulation & $L$ & $\Om$ & $\Ov$ & $\Ob$ & $h$ & $\sigma_8$ & $n_\mathrm{s}$ & $\Gamma$ \\ [0.5ex] 
\hline
\lcdm & $780\Mpc$ & 0.25 & 0.75 & 0.045 & 0.73 & 0.9 & 1 & -- \\
\tcdm & $500\Mpc$ & 1 & 0 & -- & 0.5 & 0.8 & 1 & 0.21 \\
\hline
\end{tabular}
\caption{Cosmological parameters for our simulations. As a target we
  use a \lcdm model with a WMAP1 type cosmology and as an input
  model we simulate a CDM-only model with a DEFW
  (\citealt{Davis1985}) spectrum having a similar spectral shape
  ($\Gamma=0.21$) to that of the \lcdm model but lacking a BAO
  feature. Each simulation was run with $N^3=512^3$ particles,
  gravitational forces were softened at $20\kpc$ and initial
  conditions generated using \texttt{N-GenIC} on an initial glass load
  at a starting redshift $z_{\mathrm i}=199$.}
\label{tab:simulations}
\end{table*}

\begin{table*}
\centering
\begin{tabular}{c c c c c c c c c c c c}
\hline\hline 
Original & Target & $z$ & $z'$ & $s$ & $M_1$ & $M_2$ & $s_\mathrm{m}$ & $k'_\mathrm{nl}$ & $b_\mathrm{eff}$ \\[0.5ex] 
\hline
\tcdm & \lcdm & 0.22 & 0 & 1.56 & $\sform{2.58}{13}\Msun$ & $\sform{7.19}{15}\Msun$ & 0.95 & $0.15\iMpc$ & 1.57 \\
\hline
\end{tabular}
\caption{Best fit scaling parameters $s$ and $z'$ for scaling between our original \tcdm model and our target \lcdm model together with the mass range of haloes in the original simulation ($M_1\rightarrow M_2$) and the non-linear wave number in the target cosmology $k_\mathrm{nl}$.}
\label{tab:scaling_parameters}
\end{table*}

Initially, we ran the original simulation to $z=0$ in a box of size $L$; but having computed the best scaling parameters ($s$, $z$) to match the target simulation, we then re-ran the original simulation to generate an output at exactly redshift $z$. In practice one would need to interpolate particle positions between simulation outputs near to redshift $z$ if one was interested in particles, or constrain the scaling redshift to be one of a set of $z$ (close to the best fit) for which one already had an output. This would be necessary in the case of halo catalogues because halo mergers mean that it is not straightforward to interpolate haloes between catalogues at different epochs. We also ran a simulation of the target cosmology to $z'=0$ in a box of size $L'=sL$, using the same mode phases and amplitudes for the displacement fields so that structures in the target and scaled simulations could be compared directly without the added complication of cosmic variance.

Halo catalogues were generated with the public FoF code \texttt{www-hpcc.astro.washington.edu/tools/fof.html}, using a linking length of $b=0.2$ times the mean inter-particle separation. No attempt was made to reject unbound particles.

For our simulations the best-fit scaling parameters are \new{summarised in Table \ref{tab:scaling_parameters}}. The match to the mass function produced by this scaling was shown in MP14 and is good only to the $10\%$ level across all masses, despite $\sigma(R)$ being matched to within $1\%$ across all scales relevant to these haloes (Fig. 1 of MP14). A similar level of disagreement in the measured mass function was found in AW10 (their Fig. 7) in converting between WMAP1 and WMAP3 cosmologies and plausibly reflects non-universality of the mass function.

\section{Power spectra}
\label{sec:power}

\subsection{Estimation}

In order to measure power spectra in redshift space, the particles are moved to their redshift-space positions. We use the distant-observer approximation, and move particles to redshift space positions $\mathbf{s}$, via
\begin{equation}
\mathbf{s}=\mathbf{x}+\frac{\mathbf{v}\cdot\hat{\mathbf{x}}_i}{aH(a)}\hat{\mathbf{x}}_i
\end{equation}
where $\mathbf{x}_i$ is arbitrarily chosen as one of the three coordinates and $\mathbf{v}$ is the peculiar velocity. The density field is then created using nearest grid point (NGP) binning of the particles onto an $m^3$ mesh; the effect of this in Fourier space is allowed for by dividing by the normalized transform of a cubic cell,
\begin{equation}
J(\mathbf{k})=\sinc{(\theta_x)}\, \sinc{(\theta_y)}\, \sinc{(\theta_z)}\ , 
\end{equation}
where $\theta_i=k_i L/2m$.

We compute the 2D anisotropic power spectrum of both haloes and particles by binning linearly in $\mu=\cos\theta$, where $\theta$ is the angle of the wavevector to the chosen line of sight (LOS), and logarithmically in $k$, between the box scale mode and the Nyquist mode. We use the dimensionless redshift-space power spectrum, $\Delta^2(k,\mu)$, defined as the contribution to the redshift space variance per logarithmic interval in $k$ and linear interval in $\mu$
\begin{equation}
\sigma_s^2=\int_{-1}^{1}\,\mathrm{d}\mu\int_0^\infty \,\mathrm{d}\ln k\;\Delta_\mathrm{s}^2(k,\mu)\ .
\end{equation}
Once $\Delta^2(k,\mu)$ has been computed from the particle or halo distribution we additionally compute the monopole ($\ell=0$) and quadrupole ($\ell=2$) moments of the full 2D distribution (\eg \citealt{Cole1994}). These are given by
\begin{equation}
\Delta^2_\ell(k)=\frac{2\ell+1}{2}\int_{-1}^1 P_\ell(\mu)\,\Delta_\mathrm{s}^2(k,\mu)\,\mathrm{d}\mu\ ,
\label{eq:multipoles}
\end{equation}
where $P_\ell(\mu)$ are the Legendre polynomials; $P_0(\mu)=1$ and
$P_2(\mu)=(3\mu^2-1)/2$. All odd moments vanish due to the symmetry
$\Delta_\mathrm{s}^2(k,\mu)=\Delta_\mathrm{s}^2(k,-\mu)$. Almost all the redshift-space
signal resides in the quadrupole, and we will concentrate on this. In
order to compute the multipoles of $\Delta_\mathrm{s}^2(k,\mu)$ we fit a model
`monopole + quadrupole' to $\Delta_\mathrm{s}^2(k,\mu)$. This is necessary
because this function is sampled sparsely at low $k$ due
to the finite periodic geometry of the simulation box, so that
equation (\ref{eq:multipoles}) is not well approximated by a discrete
sum. Due to the orthogonality of the multipoles it is not necessary to
fit a model with all (infinite) Legendre polynomials included. We
always plot $\Delta^2_2(k)/k^{1.5}$ linearly in this work because at
low $k$ (box modes) the quadrupole is very noisy and can be negative,
despite the Kaiser expectation.

In making estimates of power spectra from a distribution of discrete objects, it is normal to subtract shot noise:
\begin{equation}
\Delta^2(k)\rightarrow \Delta^2(k)-4\pi\left(\frac{k}{2\pi}\right)^3\frac{L^3}{N^3}\ ,
\end{equation}
where $N^3$ is the total number of objects under study. This
correction is only important at small scales, and the shot noise
correction has been shown to be accurate even for well-evolved glass
initial conditions (\citealt{Smith2003}). However, we should note that
this correction is inappropriate if we are purely concerned with
haloes, rather than the larger number of galaxies or mass particles
reconstituted from them. The shot-noise correction attempts to remove
a small-scale randomness, but the locations of the haloes have a
physical significance: they {\it are\/} the density field. In the halo
model, the small-scale nonlinear correlations are of the form of shot
noise owing to the finite numbers of haloes, softened via convolution
with the halo profile. The halo-only plots in MP14 were in error in
this respect, since shot noise was subtracted. But in practice the
effect was small and in any case does not affect the differential
comparison of models where the mass functions have been adjusted to be
identical.

\subsection{Models}

\begin{figure*}
\centering
\makebox[\textwidth][c]{
\hspace{1.5cm}\subfloat{\includegraphics[width=73mm,angle=270,trim=3cm 8.5cm 1cm 4.5cm]{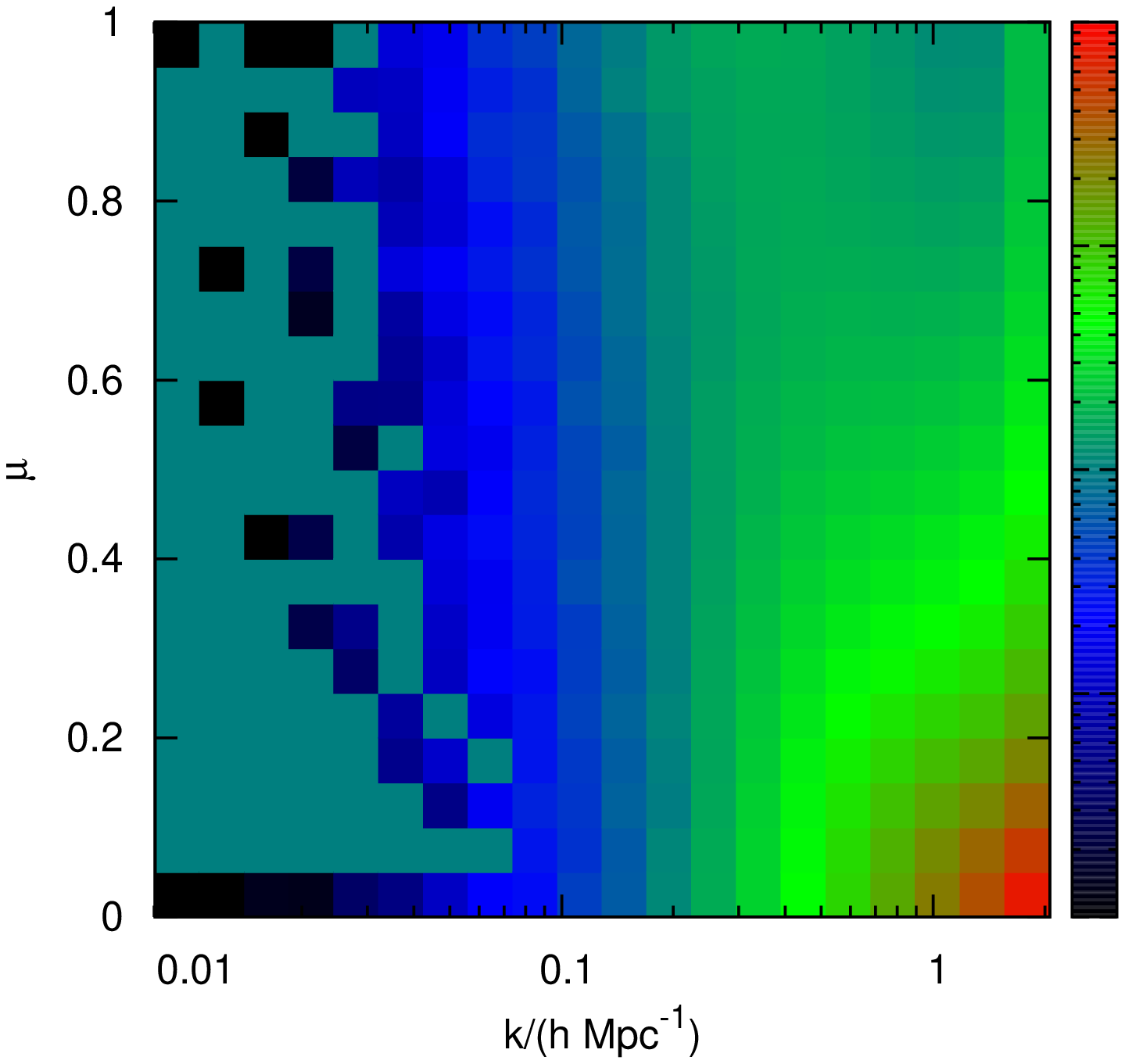}}\hspace{1cm}
\subfloat{\includegraphics[width=73mm,angle=270,trim=3cm 8.5cm 1cm 4.5cm]{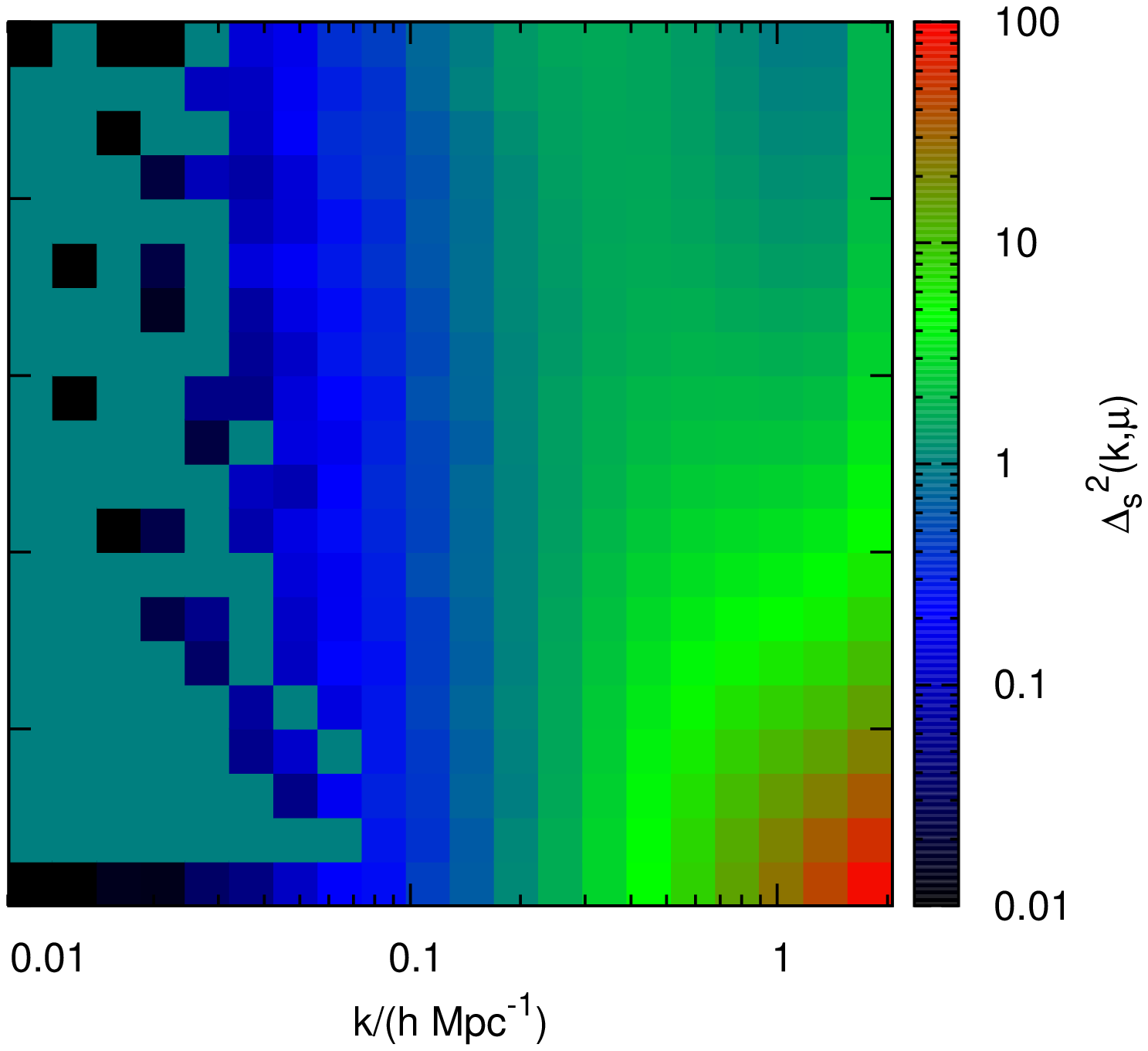}}}
\caption{The redshift-space power spectrum of the full mass distribution as a function of $k$ and $\mu$ for the rescaled simulation (left), where the AW10 method has been applied, and in addition we have restructured the internal physical and velocity structure of haloes found in the rescaled particle distribution (Section \ref{sec:restructuring}). This is compared to the spectrum measured in the target simulation (right). Differences in the spectra are difficult to identify visually and residuals are therefore shown in Fig. \ref{fig:particle_rsdpower}.}
\label{fig:particle_rsd_full}
\end{figure*}

\cite{Kaiser1987} showed that the linear theory result for the redshift space galaxy 
power spectrum, $\Delta^2_\mathrm{s,g}$, is given by
\begin{equation}
\Delta^2_\mathrm{s,g}(k,\mu)=(b+f_g\mu^2)^2\Delta^2_m(k)=(1+\beta\mu^2)^2\Delta_g^2(k)\ ,
\label{eq:kaiser_power}
\end{equation}
where $\Delta^2_m(k)$ and $\Delta^2_g(k)$ the real space matter and galaxy power spectra and $\beta=f_\mathrm{g}/b$. The growth rate features \new{in this equation} because the linear velocity field depends on the growth rate in the ZA (equation \ref{eq:velocity_scaling}); this gives a universal apparent displacement to all tracers, even though their effective real-space displacements are biased in general. Expanding $\Delta^2_\mathrm{s,g}(k,\mu)$ in terms of Legendre polynomials (\citealt{Cole1994}), the monopole and quadrupole are given by
\begin{equation}
\eqalign{
\Delta^2_0(k) &= \left(1+\frac{2}{3}\beta+\frac{1}{5}\beta^2\right)\Delta^2(k)\ , \cr
\Delta^2_2(k) &= \left(\frac{4}{3}\beta+\frac{4}{7}\beta^2\right)\Delta^2(k)\ ,
}
\label{eq:linear_rsd_spectra}
\end{equation}
and their ratio is a single function of $\beta$:
\begin{equation}
G(k)\equiv \frac{\Delta^2_2(k)}{\Delta^2_0(k)}\simeq\frac{1+\frac{2}{3}\beta+\frac{1}{5}\beta^2}{\frac{4}{3}\beta+\frac{4}{7}\beta^2}\ .
\label{eq:G_definition}
\end{equation}
It is infeasible to predict $b$ with sufficient precision to convert this to the desired growth rate, but the apparent amplitude of real-space clustering can be inferred accurately, so the pure dark-matter combination $f_g\sigma_8$ can be measured.

Equation (\ref{eq:kaiser_power}) is only valid
on linear scales and breaks down more quickly than the linear
approximation for the real-space power spectrum due to velocities
becoming non-linear before densities (\citealt{Scoccimarro2004}). More
complicated expressions for the redshift-space power exist in the
literature (\eg \citealt{Scoccimarro2004};
\citealt{Taruya2010}) and some prescriptions are compared by
\cite{delaTorre2012}. These models all attempt to deal with two
related issues: (1) deviations of the real-space power spectrum
from linear theory; (2) additional sources of anisotropy beyond the
bare Kaiser model. The latter effects amount to more rapid
damping of modes that lie close to the line of sight (\eg \citealt{Kwan2012}),
known under the generic title of `fingers-of-God' (FOG), since this
damping amounts to a radial convolution in configuration space.
A common simple model displaying both these effects is the
`Kaiser+Lorentzian' (\citealt{Peacock1994}), in which the nonlinear power spectrum is
subject to the Kaiser anisotropy and a radial damping
controlled by the pairwise dispersion $\sigma$:
\begin{equation}
\Delta^2_\mathrm{s}(k,\mu)=\frac{(b+f_g\mu^2)^2\Delta^2_{\rm NL}(k)}{1+k^2\sigma^2\mu^2/2}\ .
\label{eq:kaiser_lorentz}
\end{equation}
In practice, there are two distinct sources of FOG effects:
pure quasi-linear terms (which can be estimated e.g. via the ZA),
plus the internal velocity dispersion of haloes.
In our modelling, the first kind of FOG should be included directly,
since we work with a dynamically realistic halo catalogue.
But the internal velocity dispersions are an important contributors
to the overall effect, and we need to understand the effect on
these of internal halo rescaling.


\section{Results of rescaling}
\label{sec:results}

\subsection{Enhanced AW10 approach}

\begin{figure*}
\centering
\makebox[\textwidth][c]{
\subfloat{\includegraphics[width=60mm,trim=0cm 1.5cm 0cm 1.5cm,angle=270]{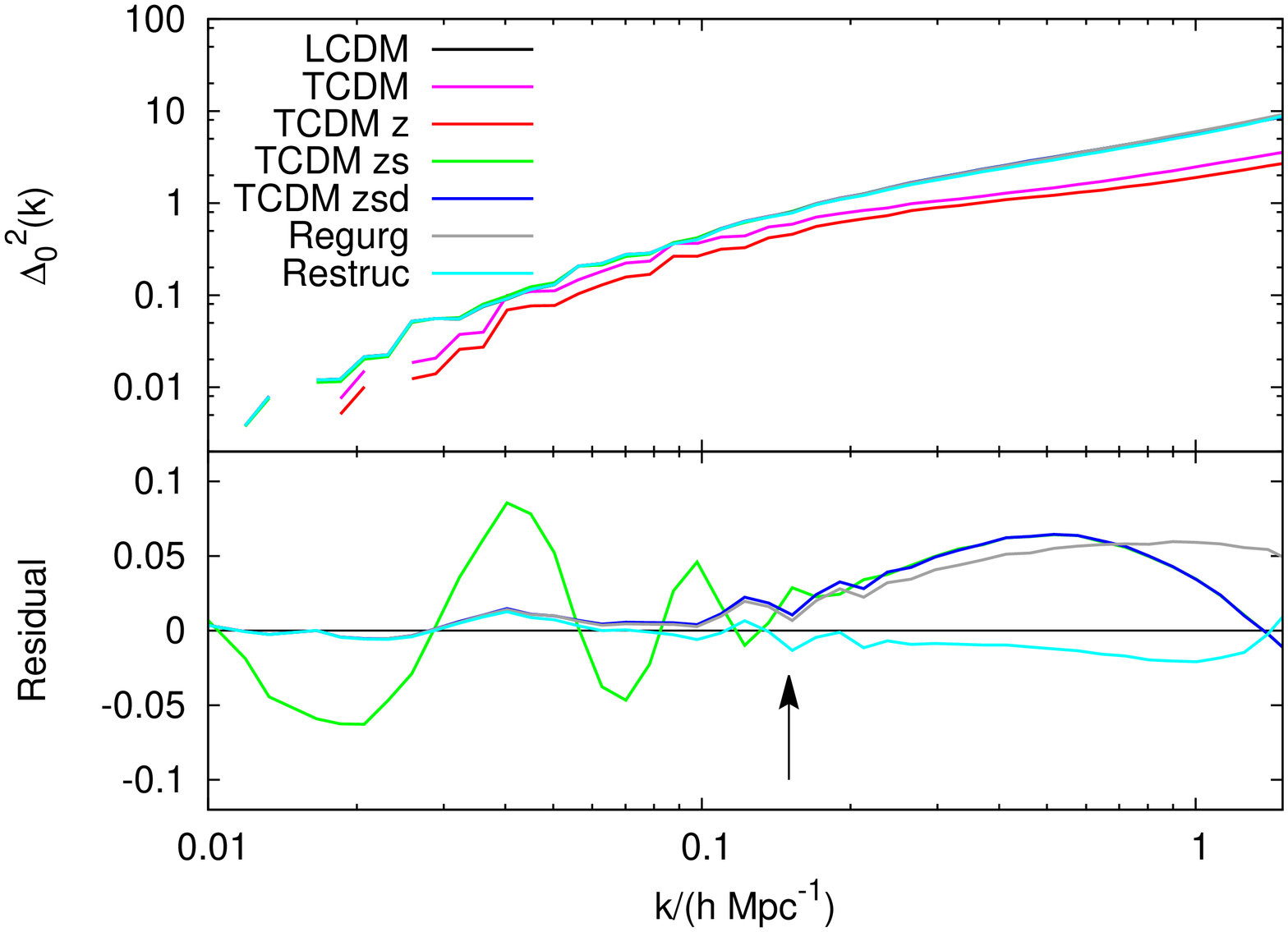}}\hspace{1cm}
\subfloat{\includegraphics[width=60mm,trim=0cm 1.5cm 0cm 1.5cm,angle=270]{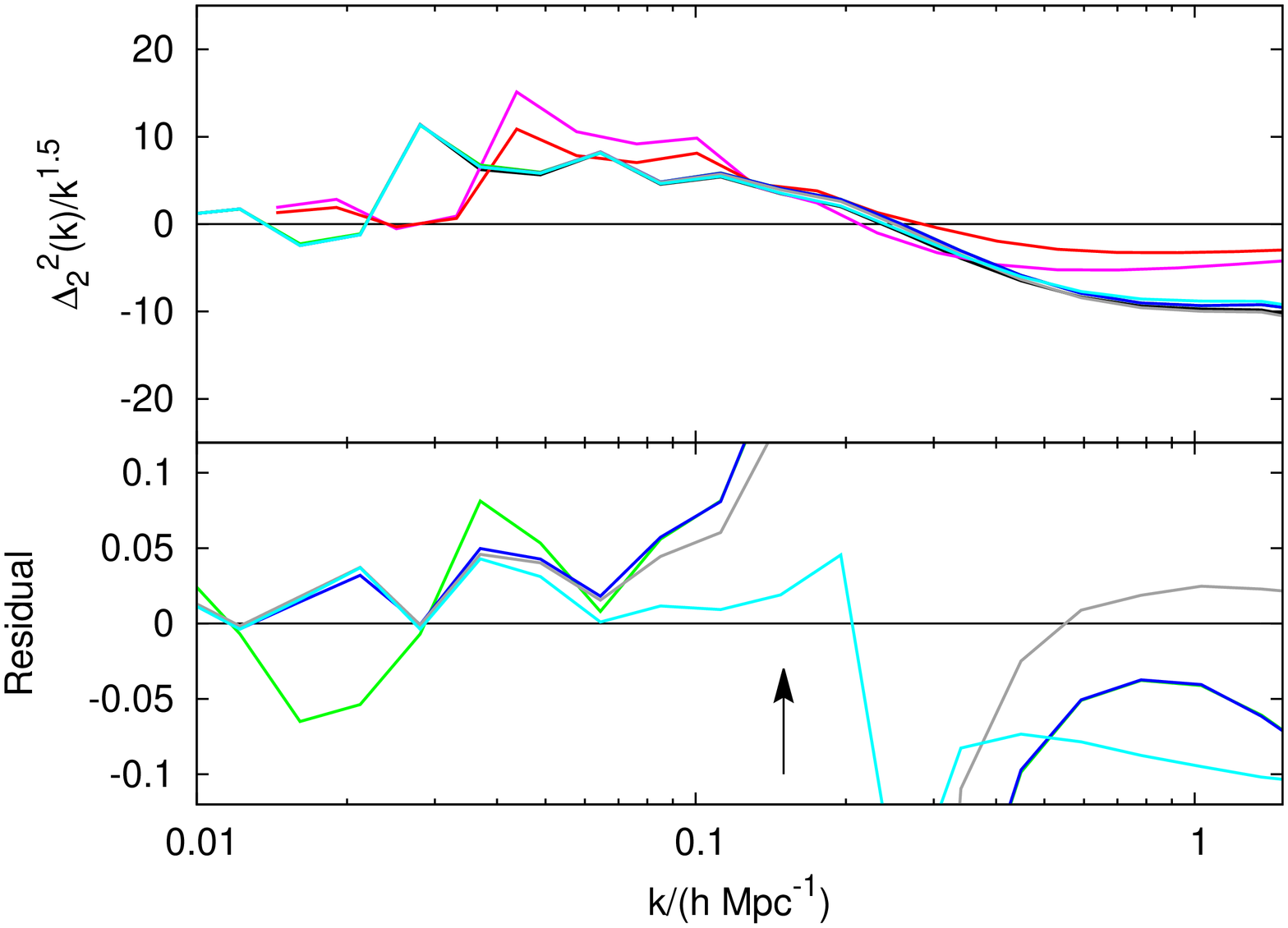}}}\\
\makebox[\textwidth][c]{
\hspace{1.5cm}\subfloat{\includegraphics[width=63mm,angle=270,trim=2cm 8cm 2cm 6cm]{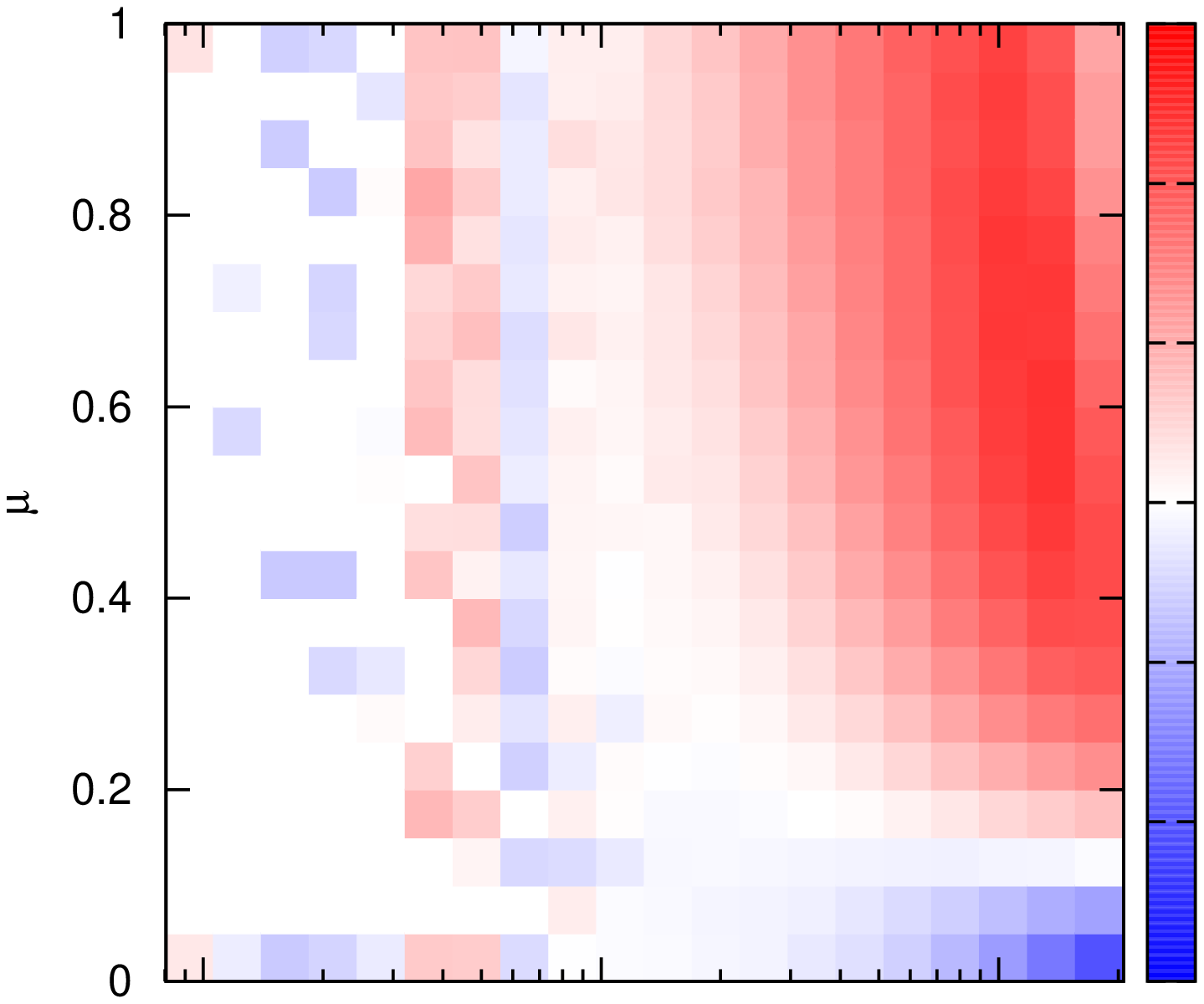}}\hspace{1cm}
\subfloat{\includegraphics[width=63mm,angle=270,trim=2cm 8cm 2cm 6cm]{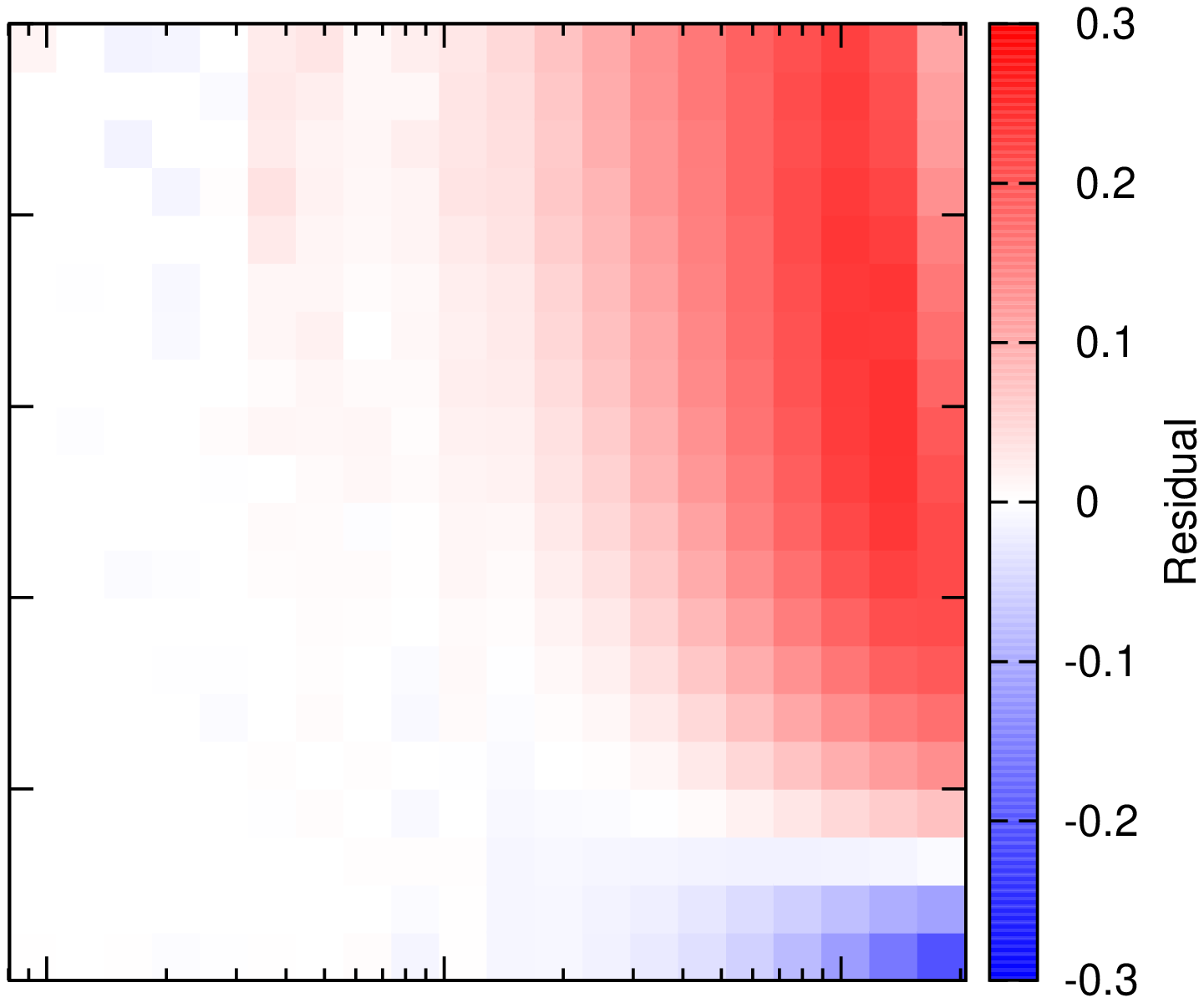}}}\\
\makebox[\textwidth][c]{
\hspace{1.5cm}\subfloat{\includegraphics[width=63mm,angle=270,trim=4cm 8cm 0cm 6cm]{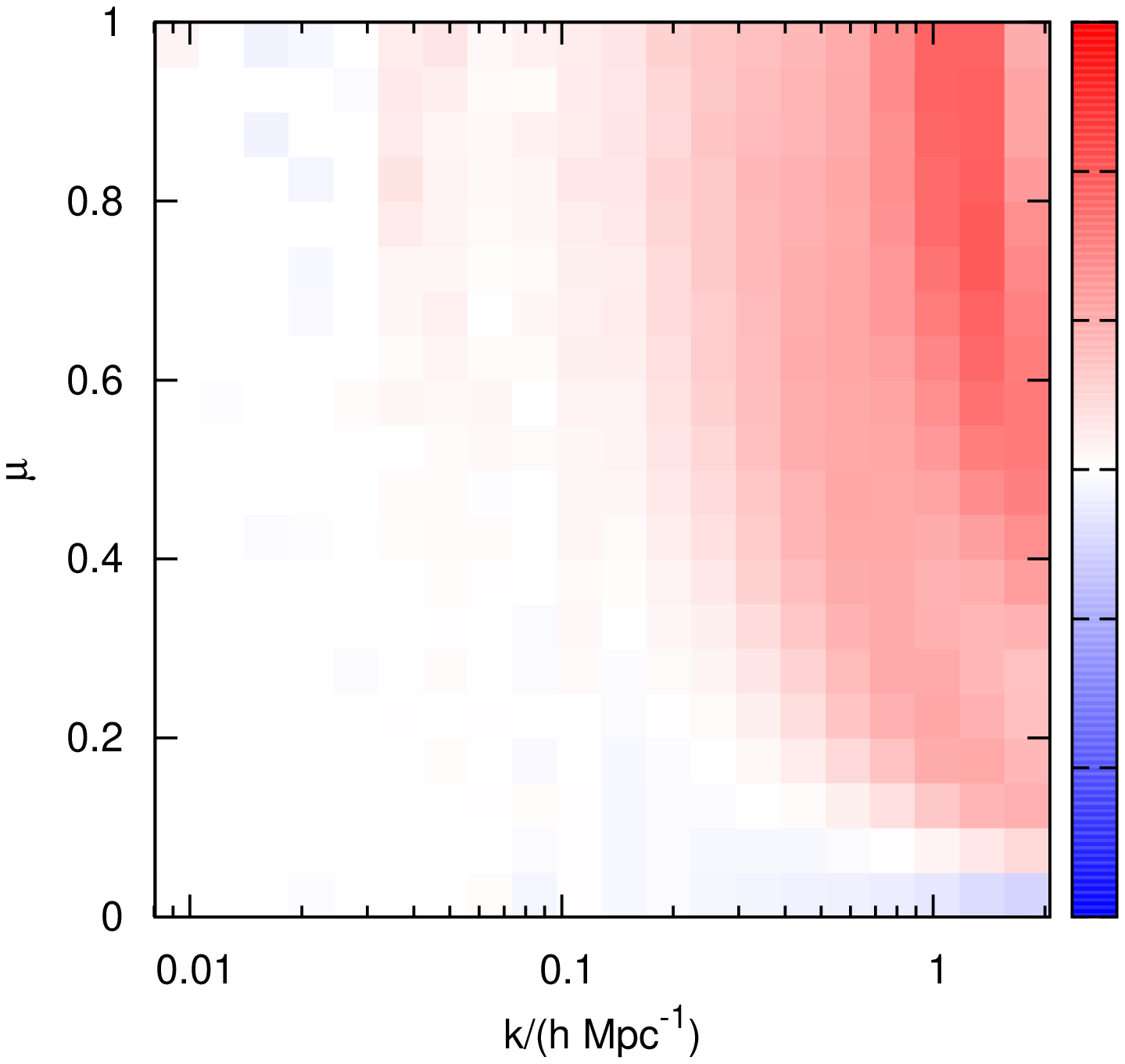}}\hspace{1cm}
\subfloat{\includegraphics[width=63mm,angle=270,trim=4cm 8cm 0cm 6cm]{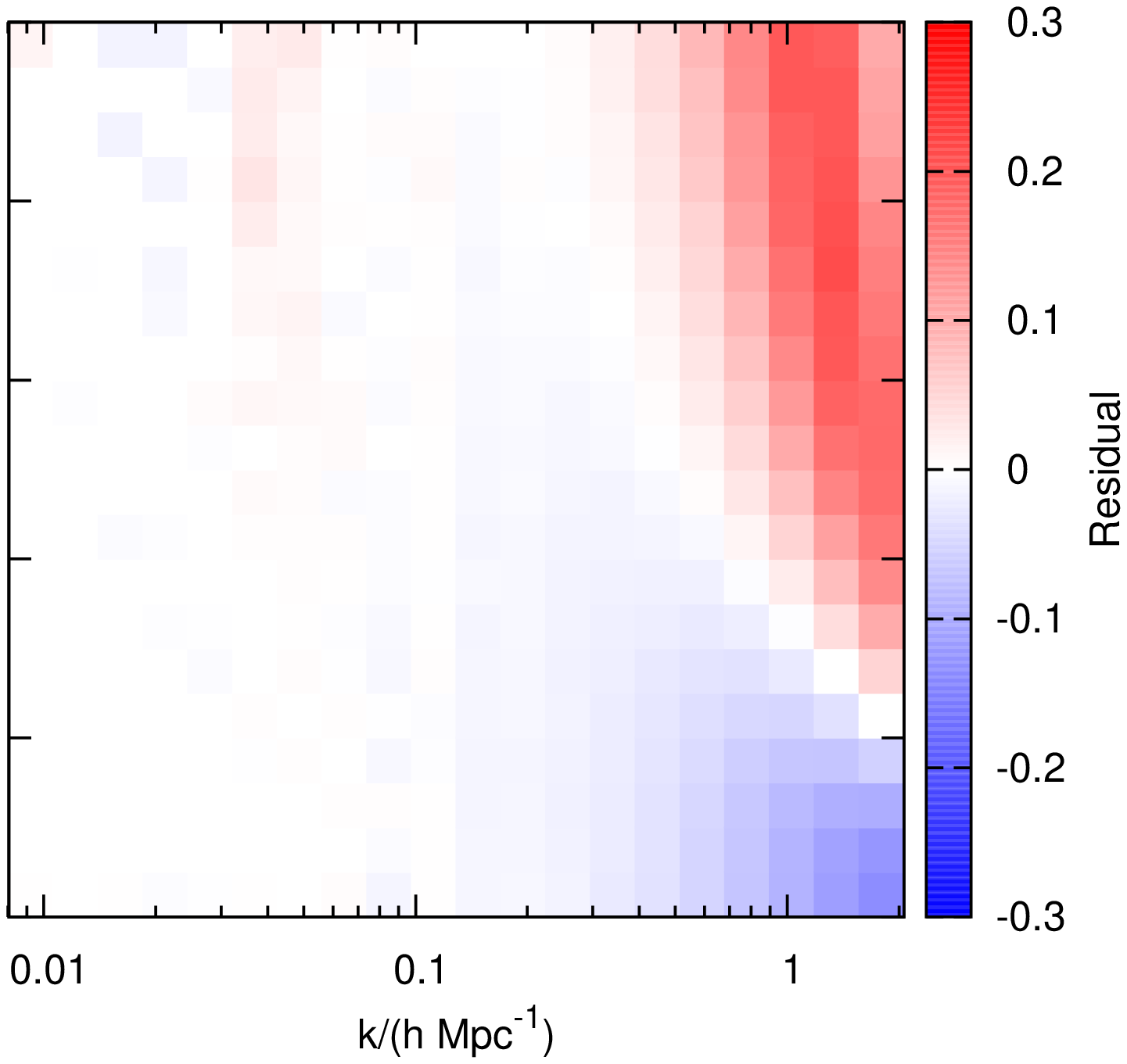}}}
\caption{Redshift-space results for a rescaled
  particle distribution, using the original AW10 method supplemented by various
alterations of halo internal structure.
  The upper two panels show the
  monopole (left) and quadrupole (right) power spectra with the arrow showing
  the non-linear scale.
  The monopole is accurate to $3\%$ to
  $k=1\iMpc$ if the restructuring method is used, including good
  reproduction of the BAO feature via
  the displacement field
  correction. The quadrupole is less well reproduced with errors at
  the $5\%$ level at large scales, but clearly applying the displacement
  field step of the method improves the match, and restructuring extends the match to
  quasi-linear scales ($k\simeq 0.2\iMpc$). At non-linear scales it is
  regurgitation that performs better. The lower four panels show the
  fractional residuals for the full redshift-space power spectrum at
  each stage of the rescaling process when compared to the target
  \lcdm simulation. The middle left panel shows the scaling in
  redshift and size and the middle right panel shows the addition of
  the displacement field step. 
  The bottom left
  panel then shows the effect of removing haloes and regurgitating
  haloes with theoretical velocity dispersions back into the
  parent particle distribution; the bottom right panel shows the
  effect of restructuring haloes. Regurgitation
  and restructuring haloes perform slightly different in the non-linear portions
  of redshift space. It is clear from this plot
  that the good agreement of the restructuring monopole is partially
  due to cancellations of errors across the full redshift-space
  plane.}
\label{fig:particle_rsdpower}
\end{figure*}

We now show a number of examples of the effect of rescaling on the
redshift-space power spectrum. We begin with the original
AW10 method applied to a full particle distribution,
with additional restructuring of internal halo properties as in
Section \ref{sec:restructuring}.
The full redshift space power spectrum before and after scaling 
is shown in Fig. \ref{fig:particle_rsd_full} where differences in 
the power between the rescaled and target simulation are difficult to see by eye.
Residuals are therefore shown in in Fig. \ref{fig:particle_rsdpower} 
together with the monopole and quadrupole moments.
The monopole is very well reproduced
(at the $1\%$ level up to $k=0.1\iMpc$), including good
recovery of the BAO feature from the large-scale displacement step,
as in the case of real space.
Deviations are seen at smaller
scales if no changes are made to the internal
structure of haloes. Improvements are gained either by restructuring
the haloes or by replacing them entirely.
To reconstitute haloes of a given mass we assigned halo
virial radii according to equation (\ref{eq:virial_radius}), took
concentrations from the \cite{Bullock2001} relations and assigned
Gaussian velocity dispersions from equation
(\ref{eq:sigma_v_basic}). Haloes were created in an aspherical manner, as
described in Section
\ref{sec:method}, although we found that the asphericity mattered very
little at the level of the redshift-space power. 

In contrast to the case of
real space, shown in Fig. 8 of MP14, restructuring haloes performs
better than regurgitation when compared in redshift space. Looking at the
ratios to the target in the full redshift-space power information,
shown in Fig. \ref{fig:particle_rsdpower}, one can see that
regurgitation performs better for perpendicular modes that are
unaffected by redshift distortions, but that restructuring
performs slightly better across all modes,
accounting for the slightly better overall prediction for the monopole.
The eventual monopole in the restructured case is good to $3\%$ up to
$k=1\iMpc$, whereas the regurgitated monopole is good to $5\%$. The
quadrupole is noisier and $5\%$ deviations can be seen up to the
non-linear scale where the fractional error blows up as the quadrupole changes
sign. However, it is still evident that the displacement field step of
the method provides a better quadrupole at large scales.

\begin{figure*}
\centering
\makebox[\textwidth][c]{
\subfloat{\includegraphics[width=60mm,trim=0cm 1.5cm 0cm 1.5cm,angle=270]{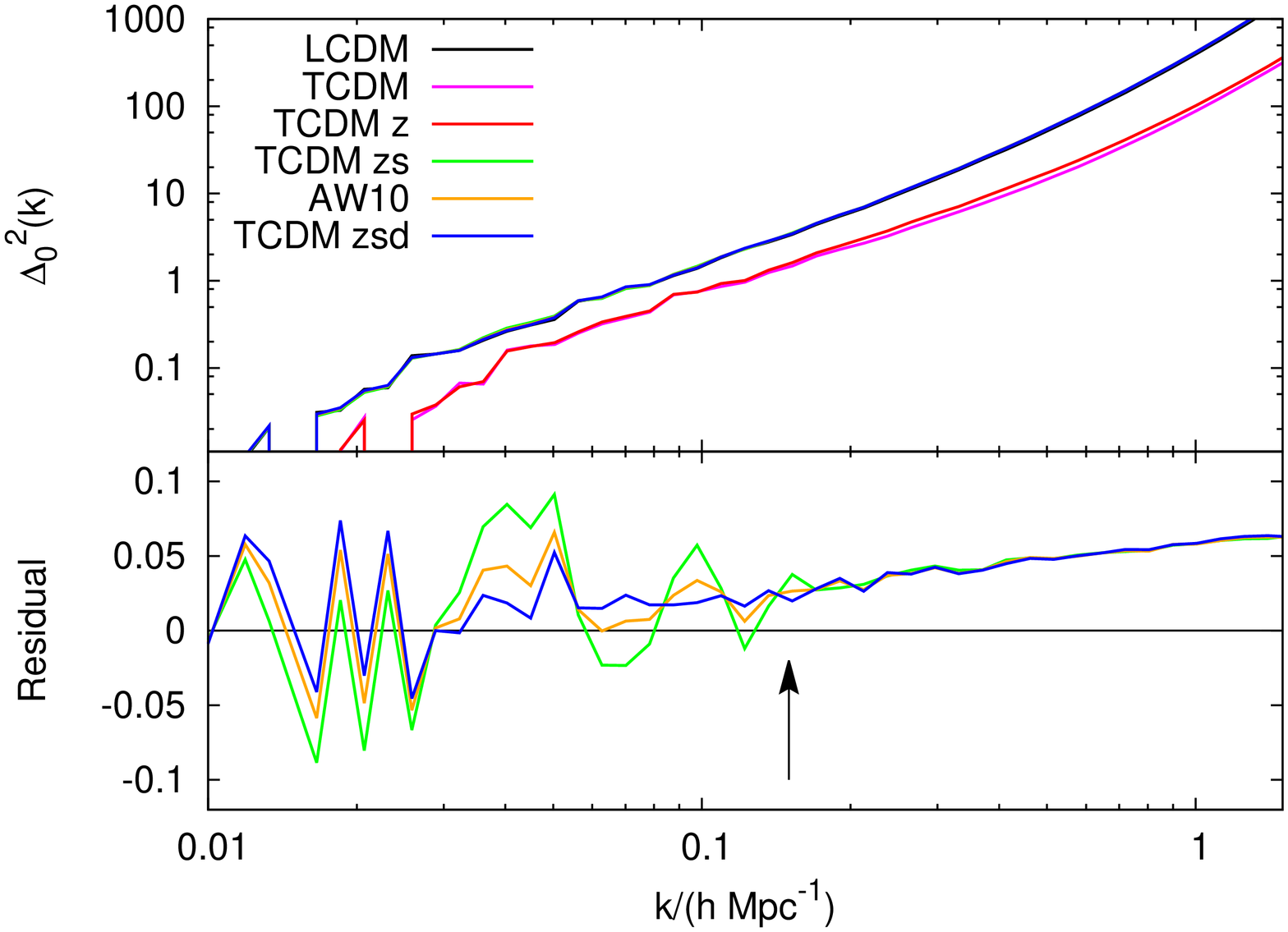}}\hspace{1cm}
\subfloat{\includegraphics[width=60mm,trim=0cm 1.5cm 0cm 1.5cm,angle=270]{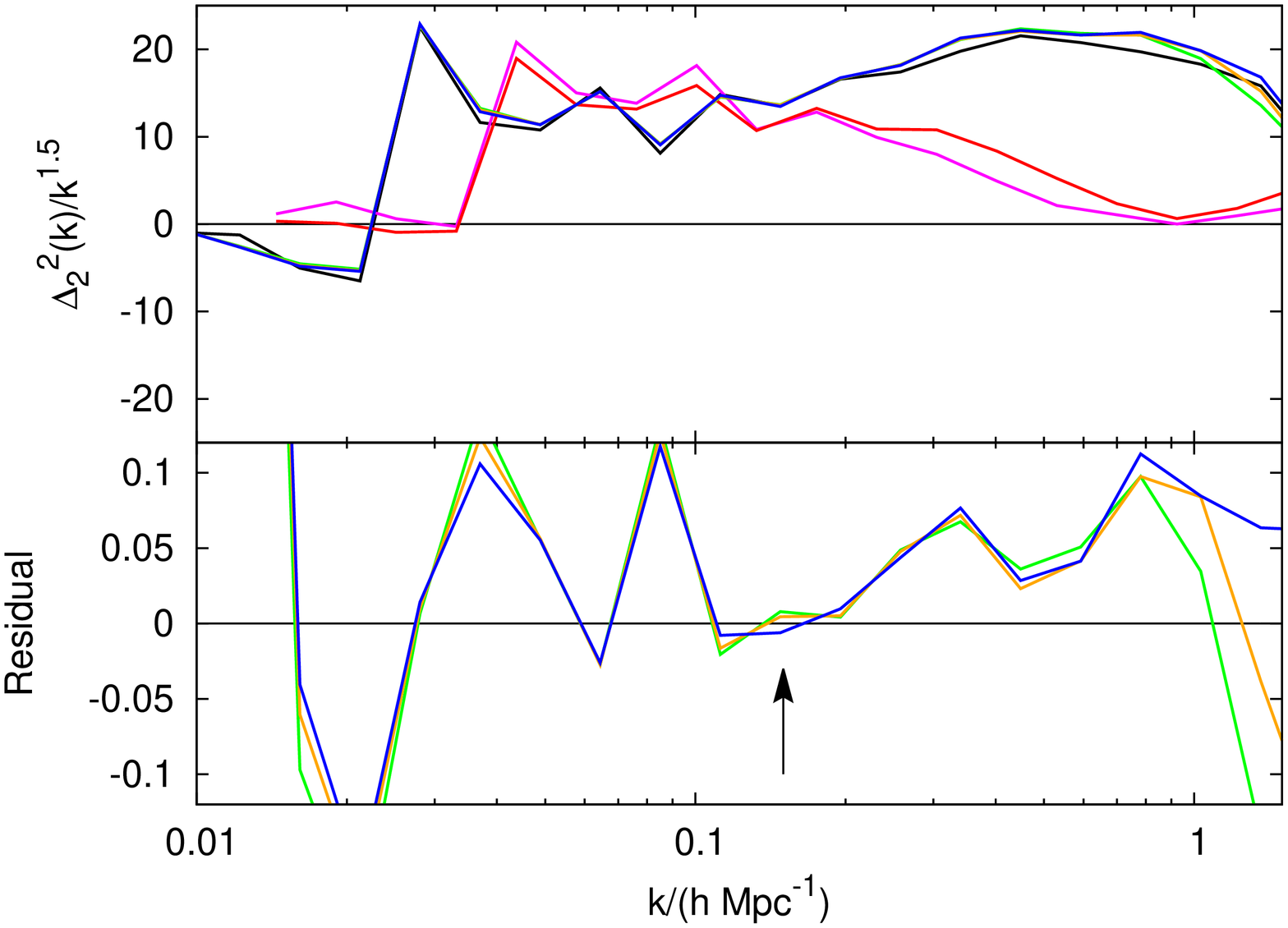}}}\\
\makebox[\textwidth][c]{
\hspace{1.5cm}\subfloat{\includegraphics[width=70mm,angle=270,trim=2.5cm 8cm 1.5cm 6cm]{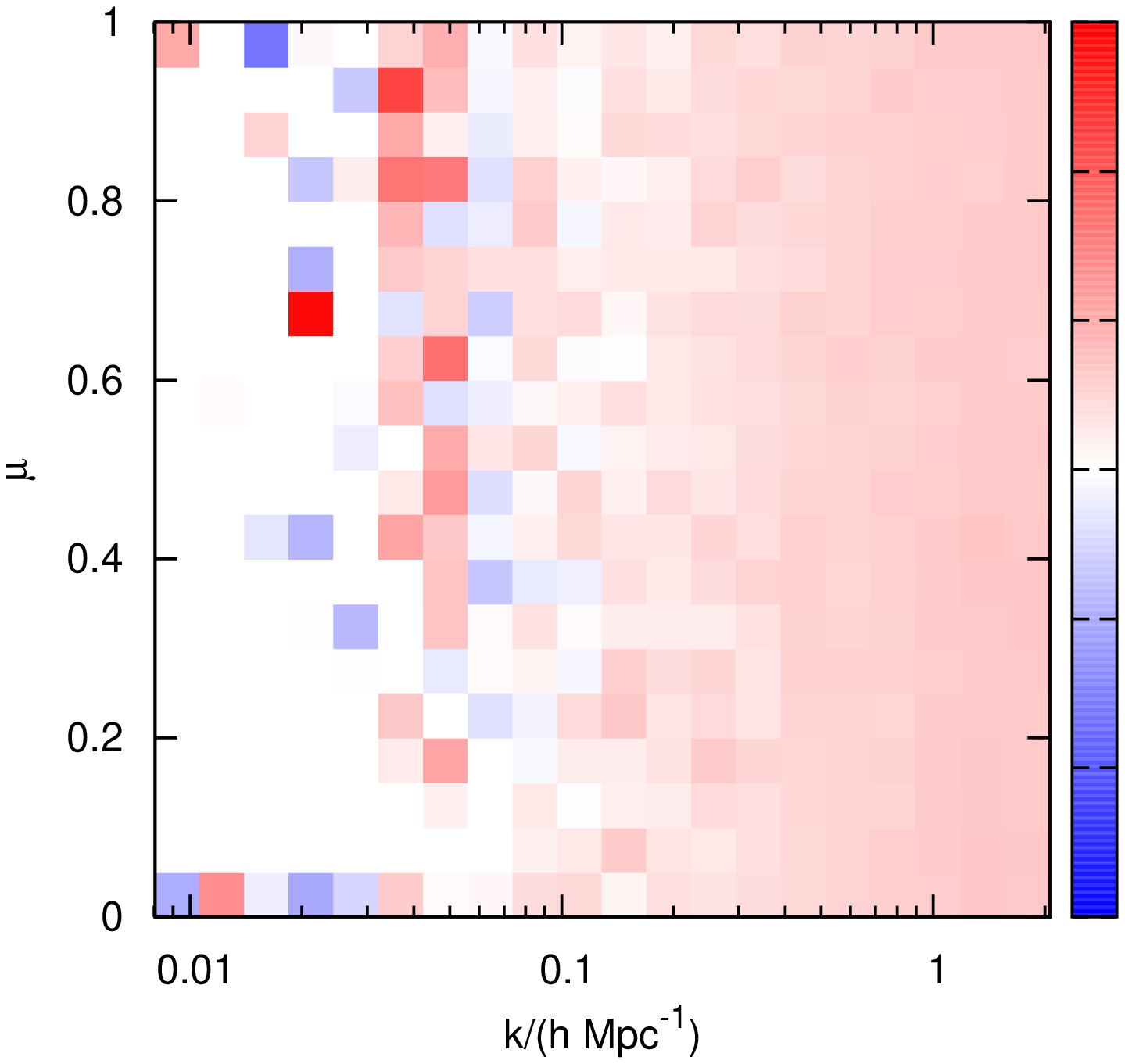}}\hspace{1cm}
\subfloat{\includegraphics[width=70mm,angle=270,trim=2.5cm 8cm 1.5cm 6cm]{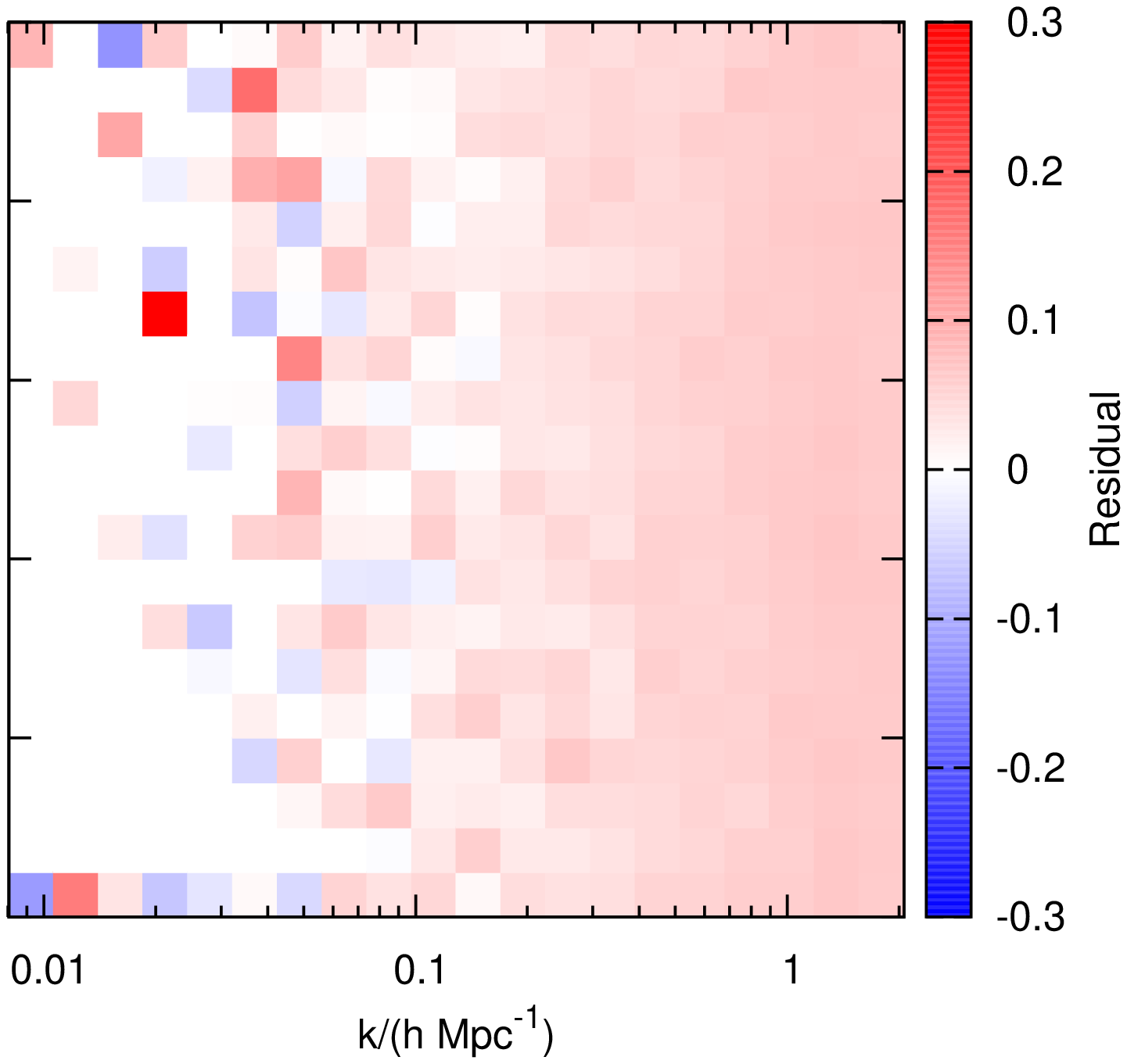}}} \\
\caption{The upper two panels show the halo redshift-space monopole
  (left) and quadrupole (right) power at each stage of the
  scaling with the arrow denoting the non-linear scale. 
  The monopole is recovered at the few $\%$ level
  across all scales, but the match surprisingly degrades somewhat at large scales. As in
  MP14 applying a biased displacement field ($zsd$) to haloes
  outperforms the original AW10 method (AW10). 
  The slant is shot noise due to differing halo numbers, which arises due to 
  the slightly imperfect match for the halo mass function. The quadrupole
  is only recovered at the $10\%$ level and contains large
  residuals that are only very slightly remedied by the displacement field correction.
  The lower panels show the fractional residuals of the 2D
  redshift-space power spectrum of the original to target simulations
  after scaling in box size and redshift (bottom left; equivalent to
  $zs$ curve) and after additionally applying a biased displacement
  field correction, and a velocity field correction, to halo positions
  (bottom right; equivalent to $zsd$ curve). The match
  is improved by applying the displacement and velocity
  field correction in the final stage of the method. The match is
  reasonable at small scales because there are no strongly non-linear
  FOG effects in the halo distribution.}
\label{fig:halo_rsdpower}
\end{figure*}

\subsection{Rescaling halo catalogues}

We now turn to the results of rescaling the halo catalogues directly.
Fig. \ref{fig:halo_rsdpower} compares the redshift-space power of the
rescaled halo distribution to that of the target haloes.
As in real space, the original AW10 method fails to
handle the BAO feature correctly until haloes are given
mass-dependent additional displacements. But the error is smaller
in redshift space, because the additional velocity field
does not depend on halo mass.
The agreement in the monopole power is around $5\%$ for
the full halo sample out to $k=1\iMpc$, but contains a slant which is shot noise due
to the numbers of haloes not being perfectly matched (recall that the
mass functions still differ at the $10\%$ level post scaling).
Note that although the non-linear scales
have not been modified here the match is still reasonable due to the lack of
an internal velocity induced FOG in the halo distribution.
However, the rescaled quadrupole is significantly less 
accurate than the monopole, with residuals at the $10\%$
level. Moreover, there is some suggestion
of a remaining large-scale residual that correlates with the BAO signal,
of amplitude about 5\% and there are some large deviations at large scales.
Additionally the displacement field step of the method seems to improve matters only marginally.

\subsection{Reconstituted haloes}

\begin{figure*}
\centering
\makebox[\textwidth][c]{
\subfloat{\includegraphics[width=60mm,trim=0cm 1.4cm 0cm 1.4cm,angle=270]{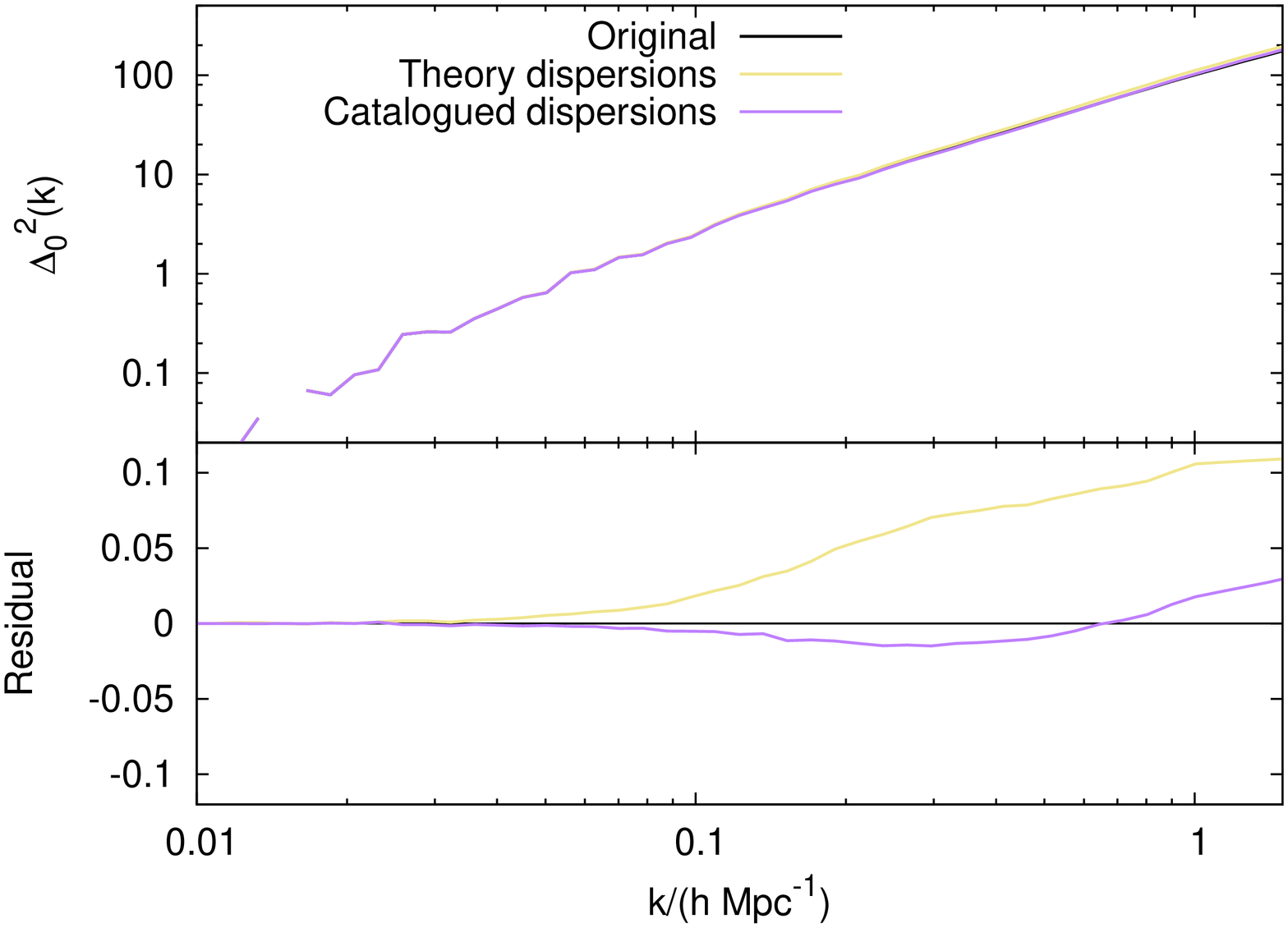}}\hspace{1cm}
\subfloat{\includegraphics[width=60mm,trim=0cm 1.4cm 0cm 1.4cm,angle=270]{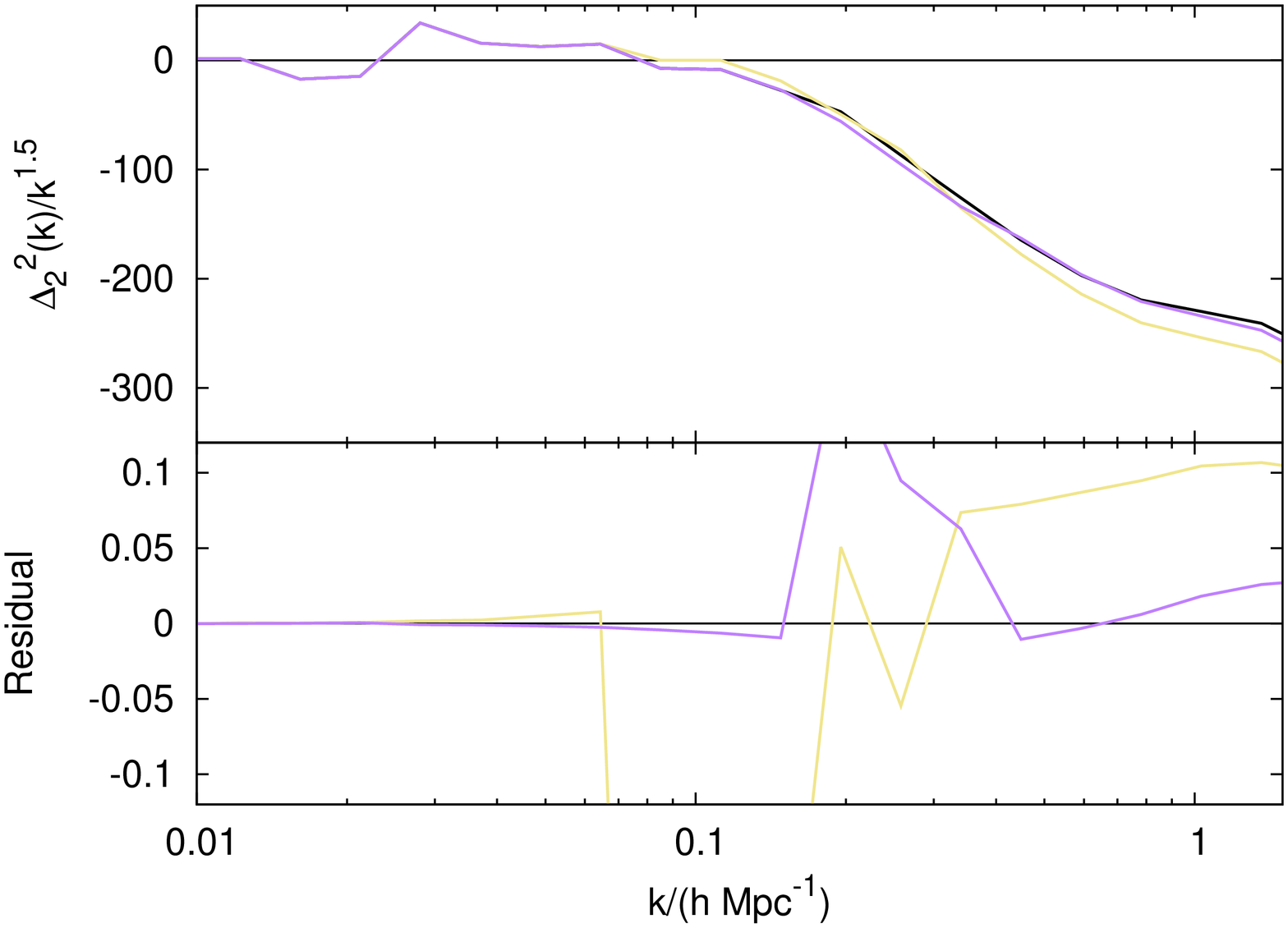}}}\\
\centering
\makebox[\textwidth][c]{
\subfloat{\hspace{1.5cm}\includegraphics[width=70mm,angle=270,trim=2.5cm 8cm 1.5cm 6cm]{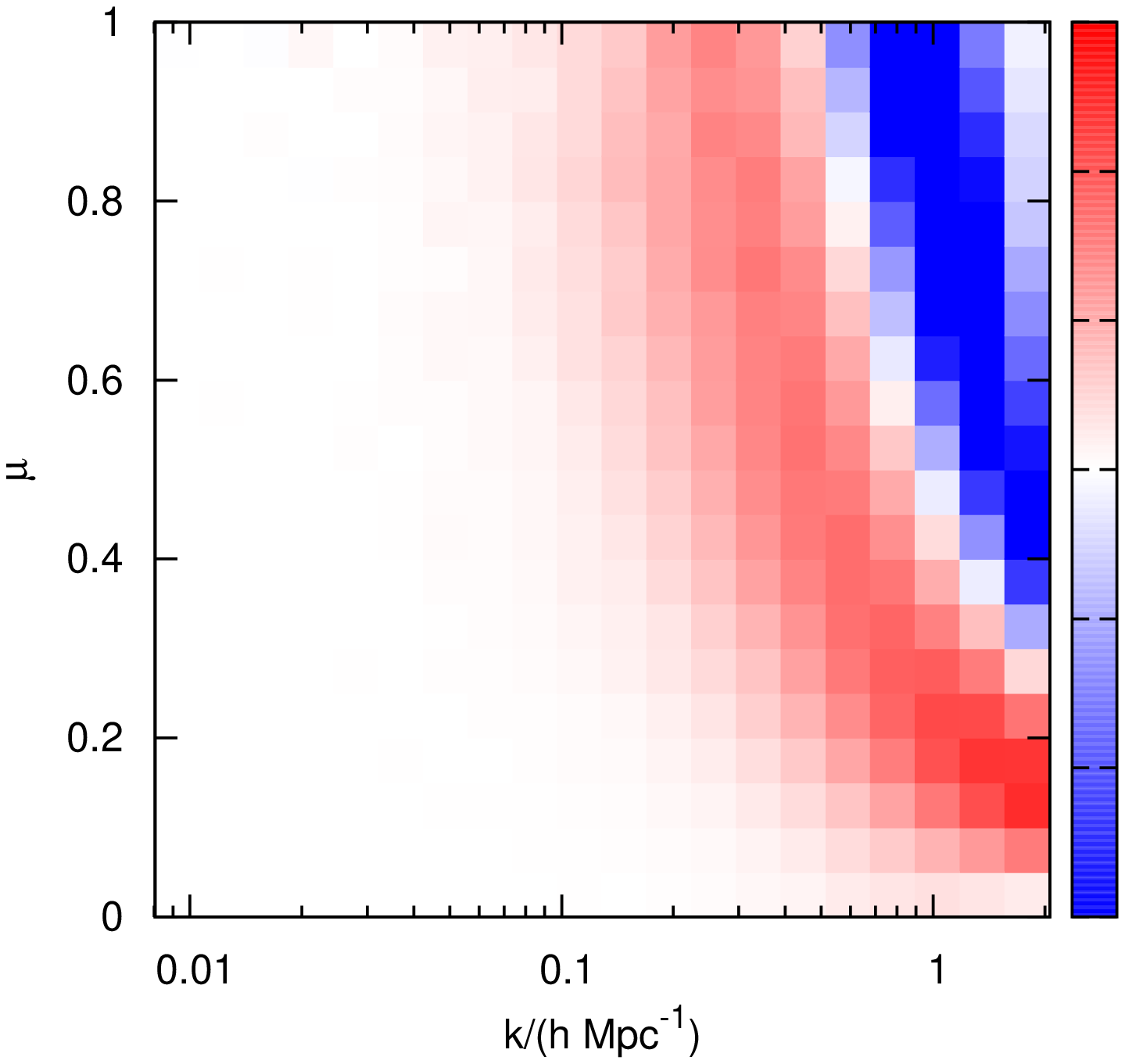}}\hspace{1.0cm}
\subfloat{\includegraphics[width=70mm,angle=270,trim=2.5cm 8cm 1.5cm 6cm]{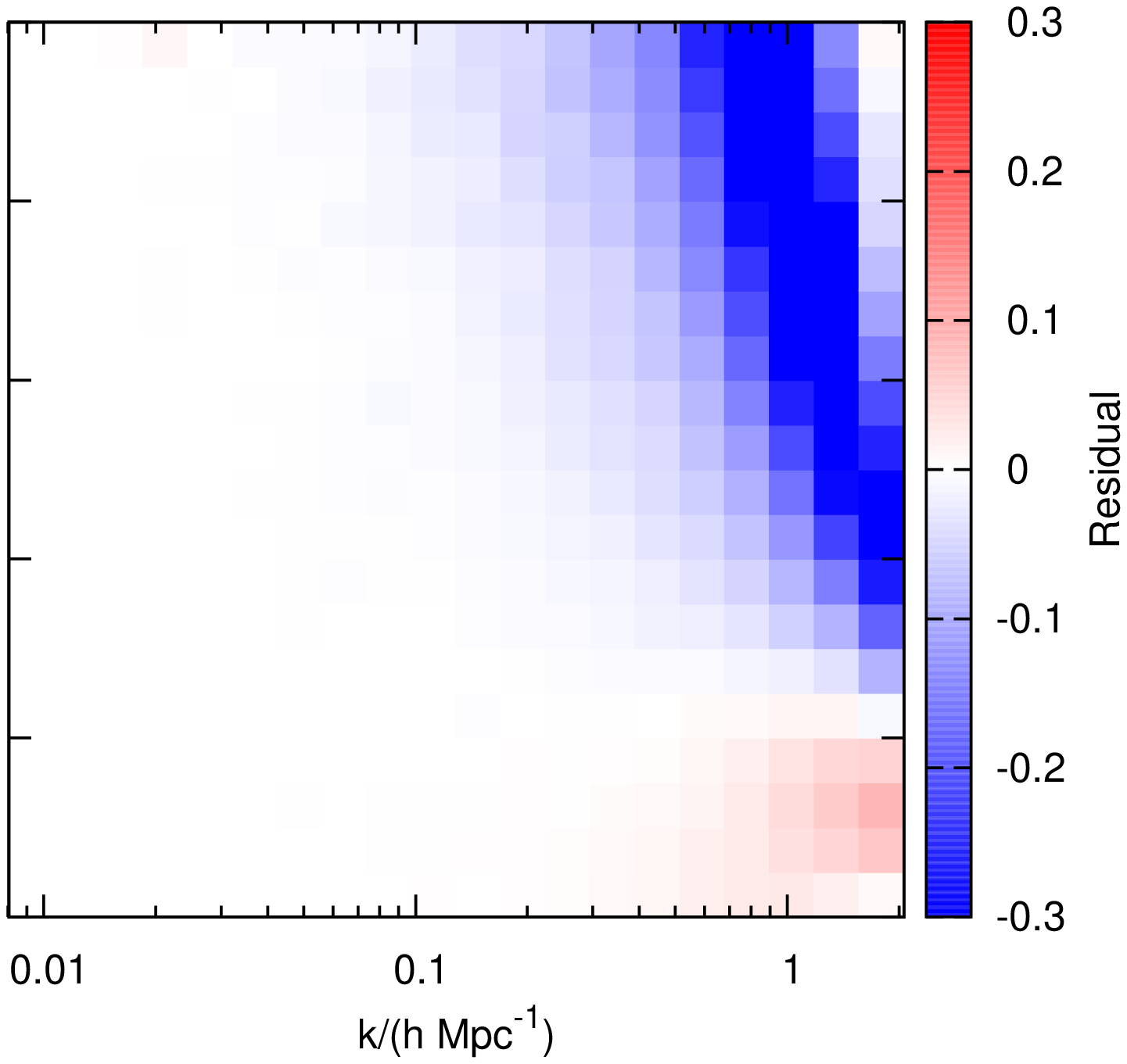}}}\\
\caption{A comparison of the power spectra of particles in haloes
  reconstituted from a halo catalogue, compared to the original
  simulation particles in the haloes. The upper panels show the
  monopole (left) and quadrupole (right) power while the lower two panels show the
  fractional residuals for the full redshift-space power spectrum as
  function of $k$ and $\mu$ in
  the case when haloes are assigned velocity dispersions using
  equation (\ref{eq:sigma_v_basic}; left; equivalent to `theory' curve above)
  as compared with the measured velocity dispersions
  (right; equivalent to `catalogue' curve above). The
  error in the quadrupole blows up around $k=0.1\iMpc$ where the
  quadrupole power changes sign. Using the catalogued
  value of the dispersions produces more accurate results, with the
  reconstituted monopole being almost perfectly reproduced (within $\simeq 1\%$)
  up to $k=1\iMpc$. The lower panels show that the error
  from using the predicted dispersion differs across
  redshift space and extends to larger scales than that from using the
  catalogued dispersions. The monopole from using catalogued
  dispersions is recovered well because errors here are concentrated
  in the high-$\mu$-high-$k$ region of the plane where the absolute
  magnitude of the power is low due to FOG. This means that the
  monopole is relatively unaffected but this is less true of the
  quadrupole, which differences in $\mu$. The match is perfect at large scales in all cases
  because the haloes are placed perfectly accurately by construction.}
\label{fig:perfect_recon_halo_rsdpower}
\end{figure*}

\begin{figure*}
\centering
\makebox[\textwidth][c]{
\subfloat{\includegraphics[width=60mm,trim=0cm 1.5cm 0cm 1.5cm,angle=270]{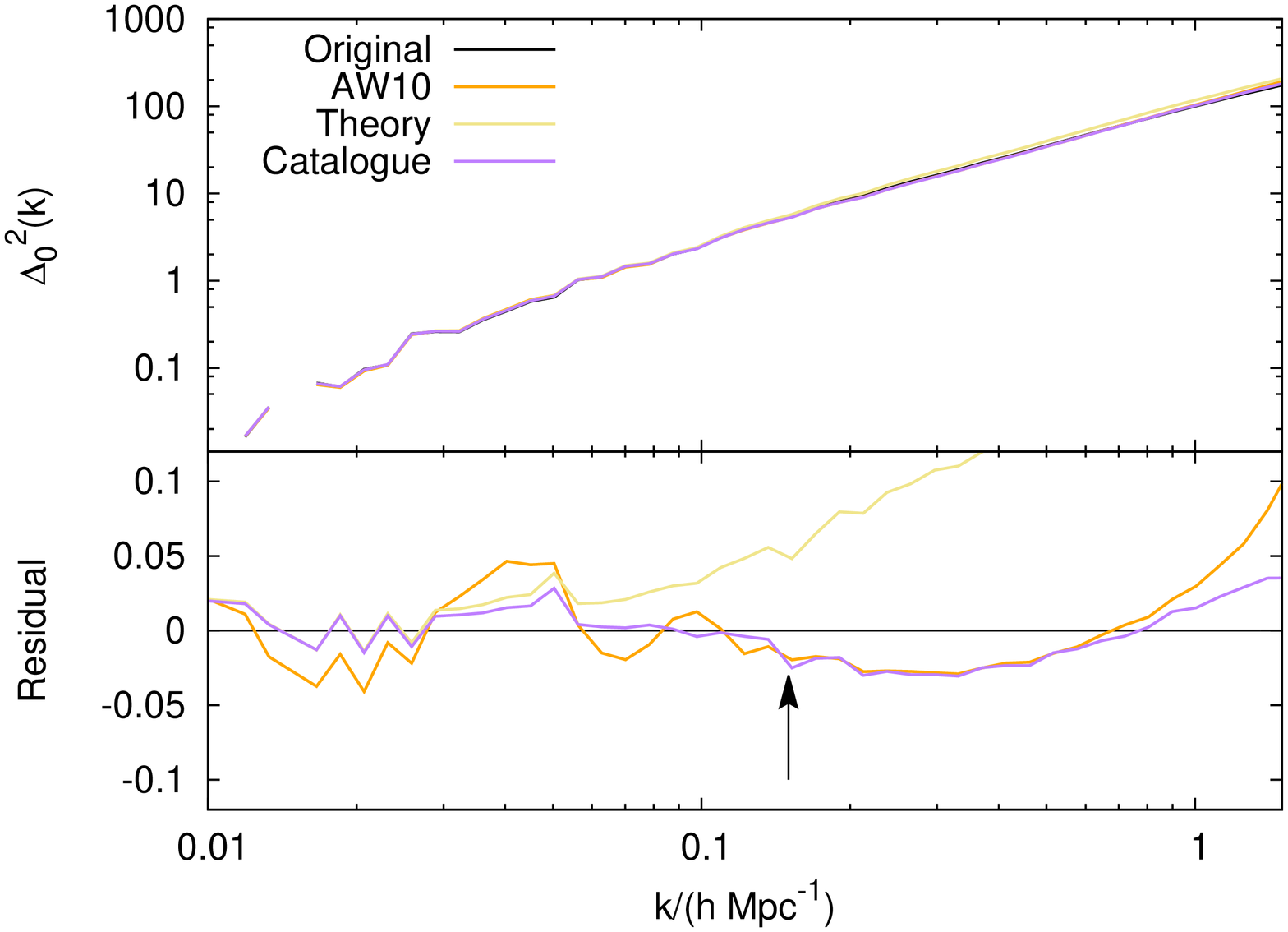}}\hspace{1cm}
\subfloat{\includegraphics[width=60mm,trim=0cm 1.5cm 0cm 1.5cm,angle=270]{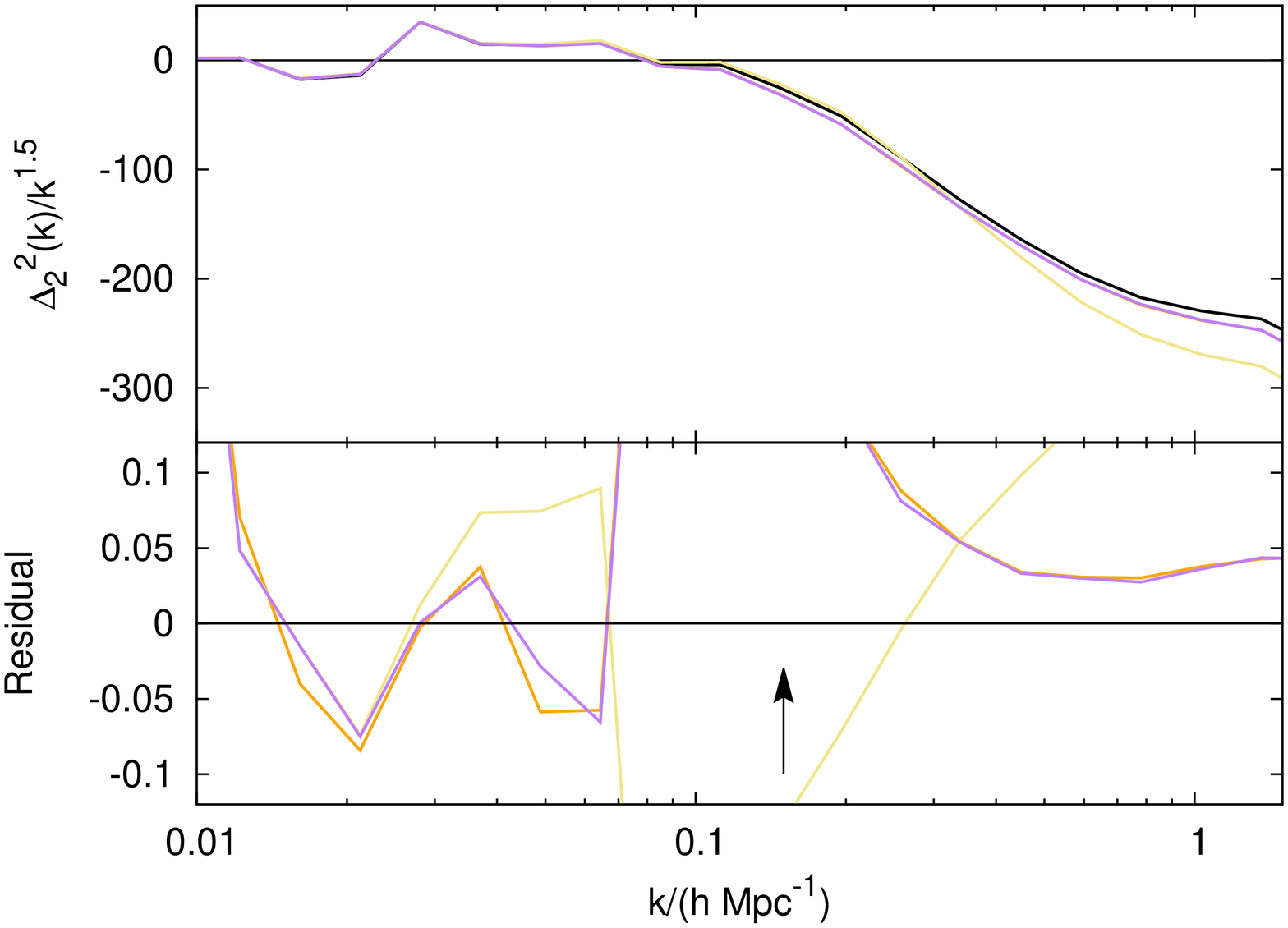}}}\\
\makebox[\textwidth][c]{
\hspace{1.5cm}\subfloat{\includegraphics[width=70mm,angle=270,trim=2.5cm 8cm 1.5cm 6cm]{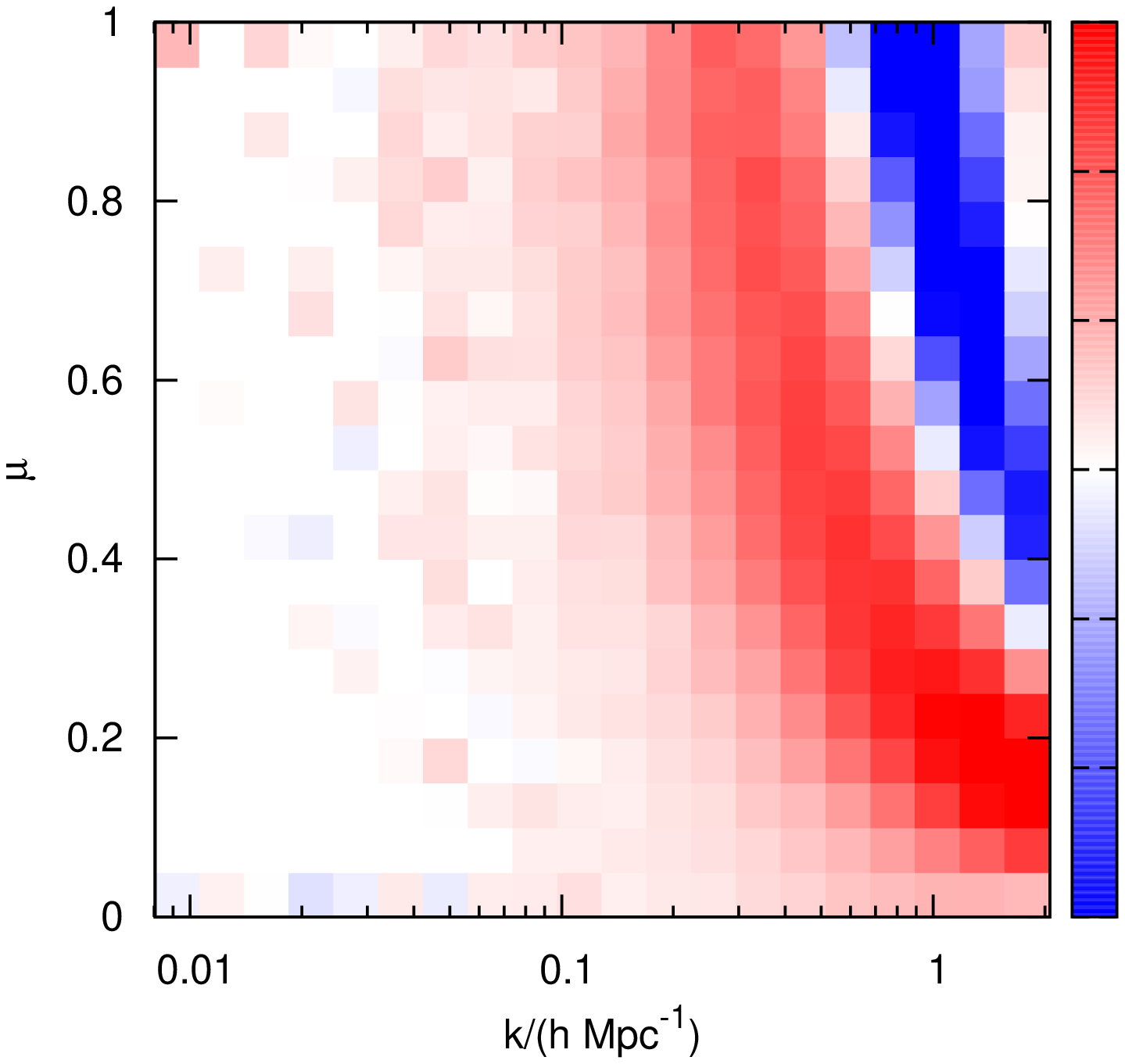}}\hspace{1cm}
\subfloat{\includegraphics[width=70mm,angle=270,trim=2.5cm 8cm 1.5cm 6cm]{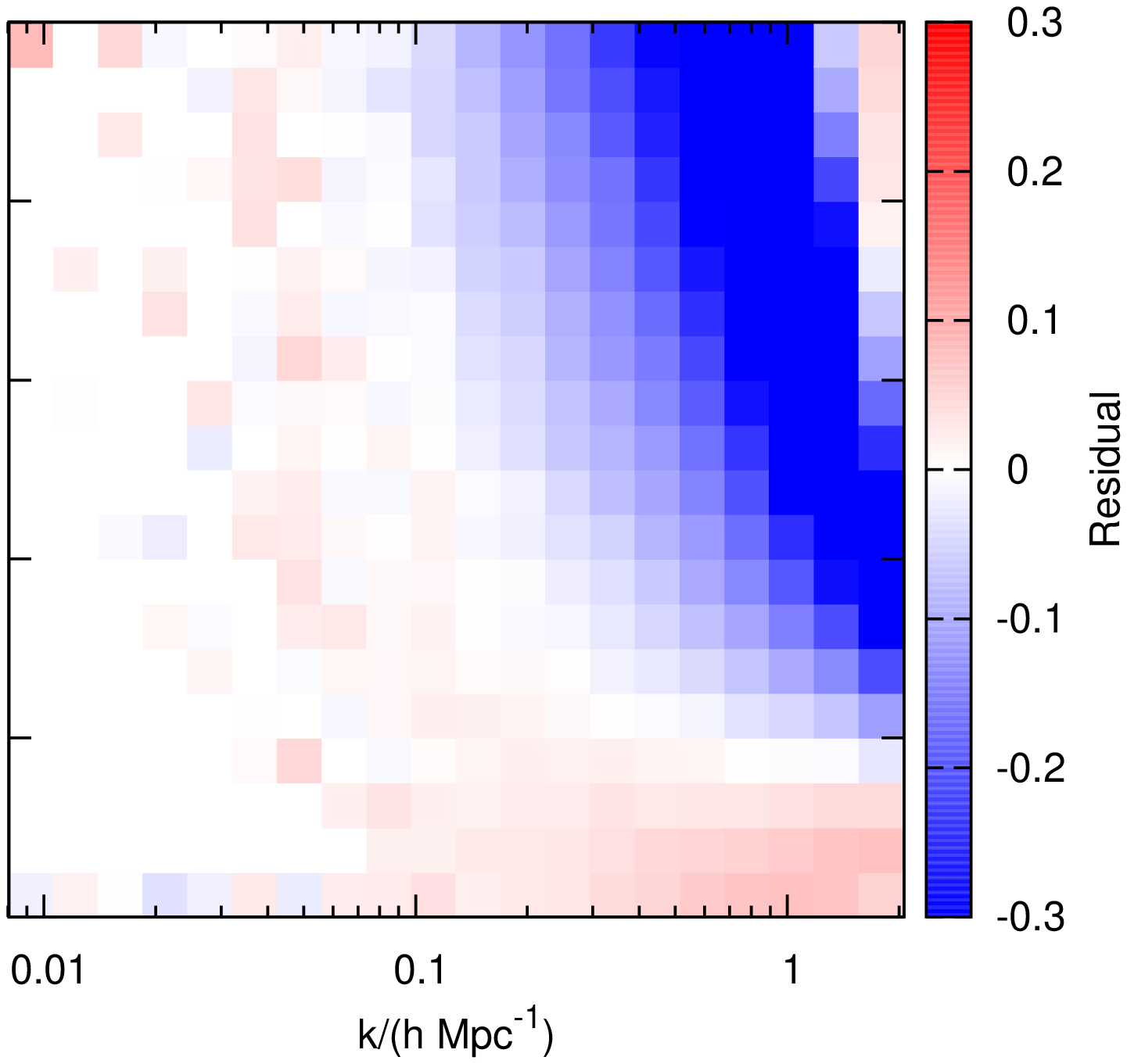}}} \\
\caption{The upper two panels show the monopole (left) and quadrupole
  (right) power spectra of particles in reconstituted haloes compared
  to particles in haloes in the target simulation. If the original
  AW10 method is used (AW10) and haloes are reconstituted with
  catalogued halo velocity dispersions a large residual BAO feature is
  seen up to the non-linear scale (arrow), which is removed if a 
  biased displacement field is used as in
  the other two curves. In these two cases better results are obtained
  for the monopole and quadrupole if rescaled catalogued dispersions
  are used (catalogue) compared to theoretical dispersions
  (theory). However, the quadrupole is relatively insensitive to the
  method used in the scaling and shows a larger error which 
  remains at the $5\%$ level on linear scales. The fractional
  residuals of the full redshift-space power spectrum is shown in the
  lower two panels from using biased displacements together with
  theoretical velocity dispersions (left, equivalent to `theory' curves above)
  and rescaled catalogued dispersions (right, equivalent to `catalogue'
  curves above). Errors in the catalogued case are concentrated at high $k$
  and high $\mu$, where there is less overall power, resulting in a
  better recovery of the monopole and quadrupole.}
\label{fig:recon_halo_rsdpower}
\end{figure*}

We now show the results of reconstituting haloes by modifying the positions and velocities of the halo particles. It is illuminating to test this aspect of our approach directly in an unscaled simulation (the $L=780\Mpc$ \lcdm simulation discussed in Section \ref{sec:simulations}). We show the redshift-space monopole, quadrupole and the full 2D redshift-space power spectrum in Fig. \ref{fig:perfect_recon_halo_rsdpower}. In both figures we compare results using reconstituted halo particles to the `exact' results based on the particles from which the halo catalogue was constructed. This tests our assumptions about both the halo internal physical and velocity structure. Haloes were reconstructed with NFW profiles (equation \ref{eq:nfw}) with concentrations from \cite{Bullock2001} and velocity dispersion was assumed to be Gaussian with a standard deviation given by equation (\ref{eq:sigma_v_basic}). The agreement in the monopole is almost perfect (within $1\%$ up to $k=1\iMpc$) if catalogued velocity dispersions are used, but deviations occur when using a theoretically predicted $\sigma_v$, which predicts too high a power, suggesting that $\sigma_v$ is too low and FOG are not strong enough. This can be seen via a direct comparison of predicted and catalogued dispersions, which show the predicted value to be low by a factor of $1.07$. According to equation (\ref{eq:virial_radius}), such a factor would arise if the virialized overdensity threshold was raised to about $\Delta_\mathrm{v}\simeq 300$. \new{We note that an increased $\Delta_\mathrm{v}$ is consistent with spherical model predictions for $\Lambda$CDM models (\eg \citealt{Bryan1998}) but is incompatible with our choice of $b=0.2$ as a linking length. We have not pursued the possibility of changing $\Delta_\mathrm{v}$} as we do not want to introduce free parameters that may have an unknown dependence on cosmology or simulation resolution (\eg \citealt{Power2006}; \citealt{Smith2014}).

Finally in Fig. \ref{fig:recon_halo_rsdpower} the redshift-space power
of particles in reconstituted haloes is compared to that of
particles in haloes in the target cosmology. The
agreement is good at linear scales once a biased displacement field
has been used. At non-linear scales the agreement is less good if one
uses the theoretical dispersion relation in equation
(\ref{eq:sigma_v_basic}) compared to using a rescaled version of the
catalogued halo velocity dispersion; in the later case the eventual agreement for the
monopole is impressive and is at the $3\%$ level across all scales
shown. Although the quadrupole has a larger error, this still remains
within $5\%$ in regions where it is not passing through zero.
In 2D redshift space
one can see that large fractional errors in the catalogued dispersion case are
concentrated at high $k$ and high $\mu$; these contribute less to the
monopole because the full power is damped here and the monopole is a
simple average. Using theoretical dispersions perform worse across the
entirety of redshift space, even at surprisingly large scales for high $\mu$. This
plot, combined with the tests in
Fig. \ref{fig:perfect_recon_halo_rsdpower} gives hope that the
theoretical dispersion relation could be modified slightly to produce
better results. We have not pursued this here as our aim
was to see how far one could get without resorting to tuning
additional parameters.

\subsection{Recovery of the growth rate}
\label{sec:growth}

\begin{figure*}
\centering
\includegraphics[width=60mm,angle=270]{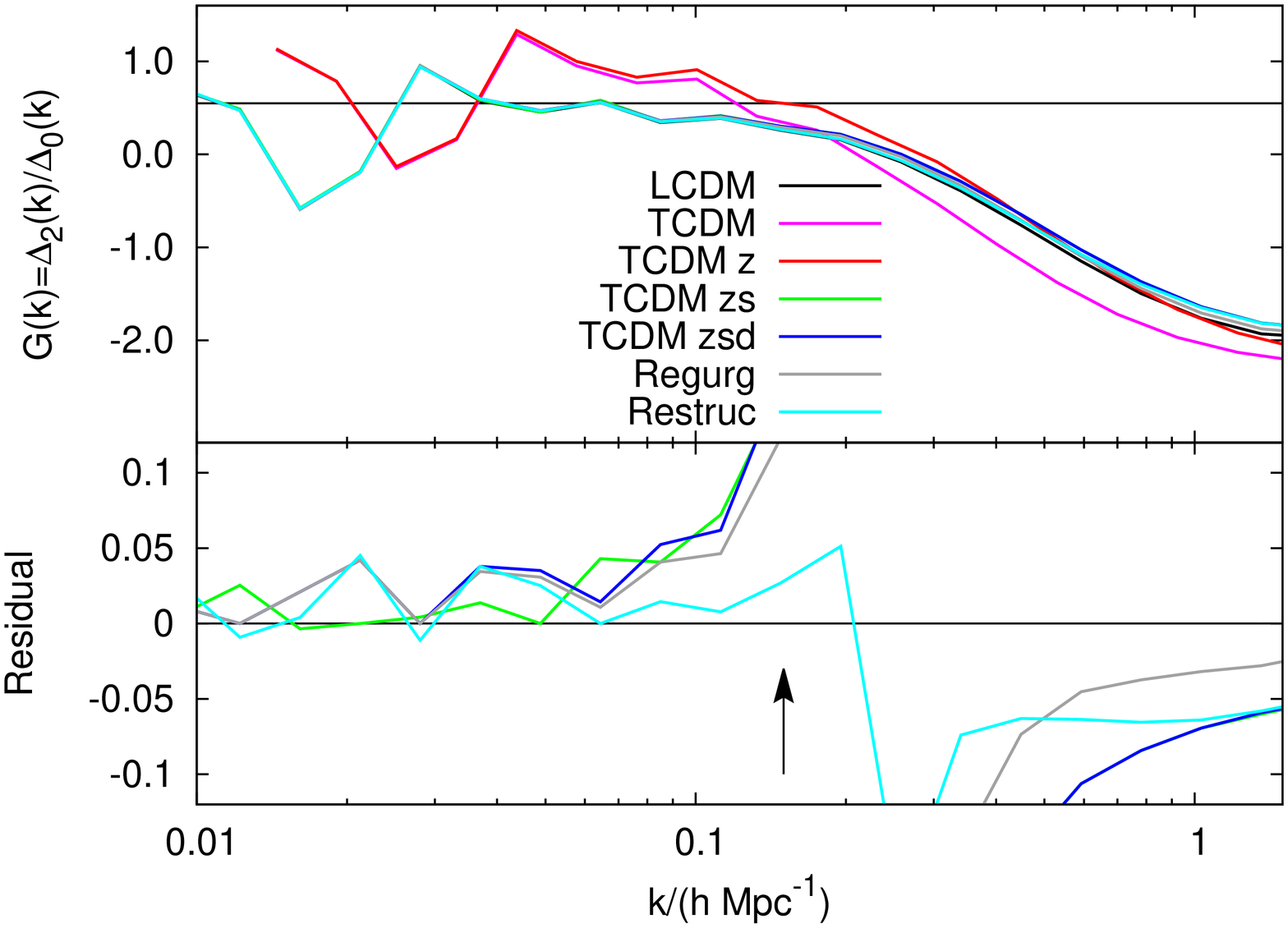}
\includegraphics[width=60mm,angle=270]{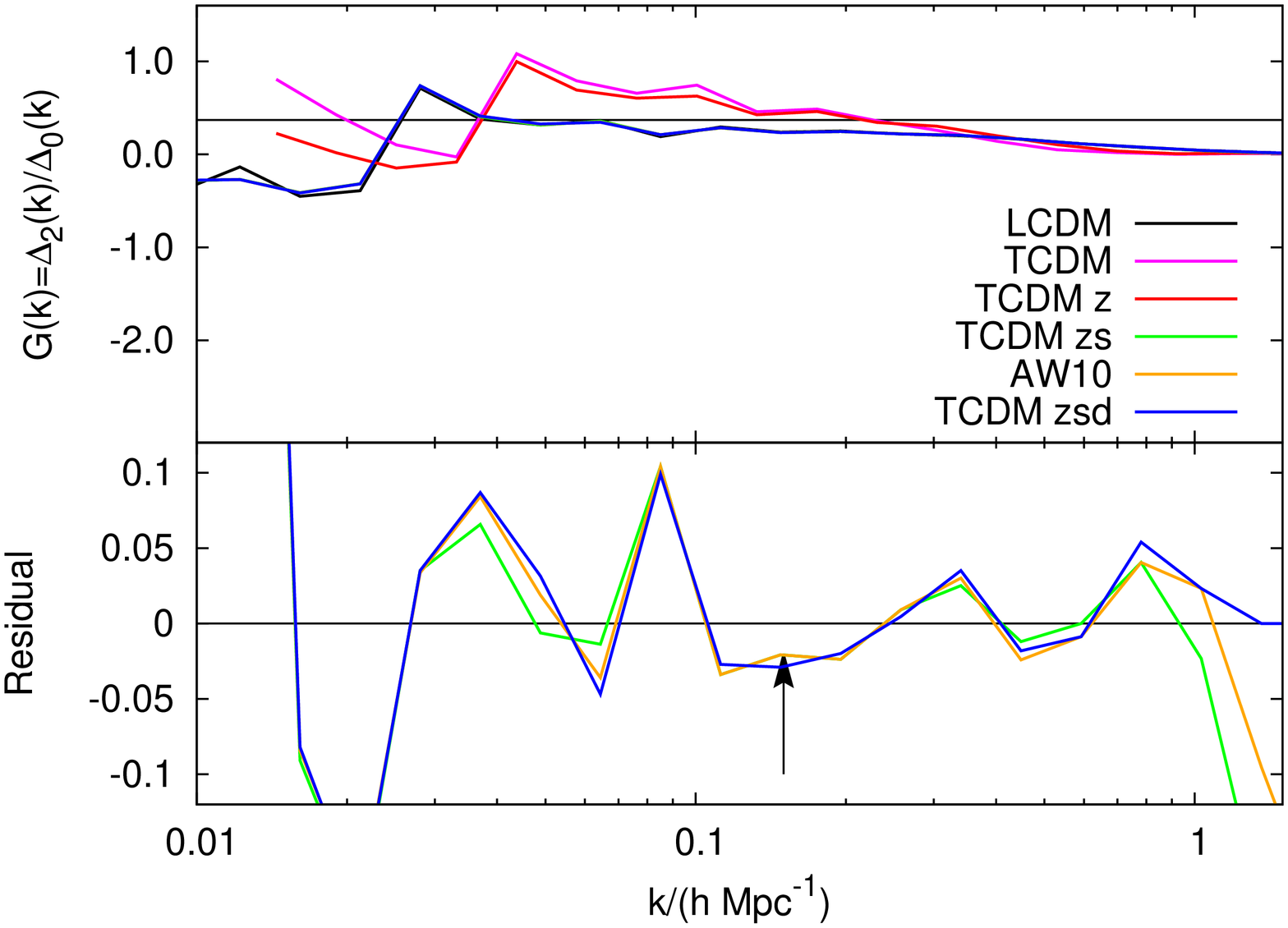}
\includegraphics[width=60mm,angle=270]{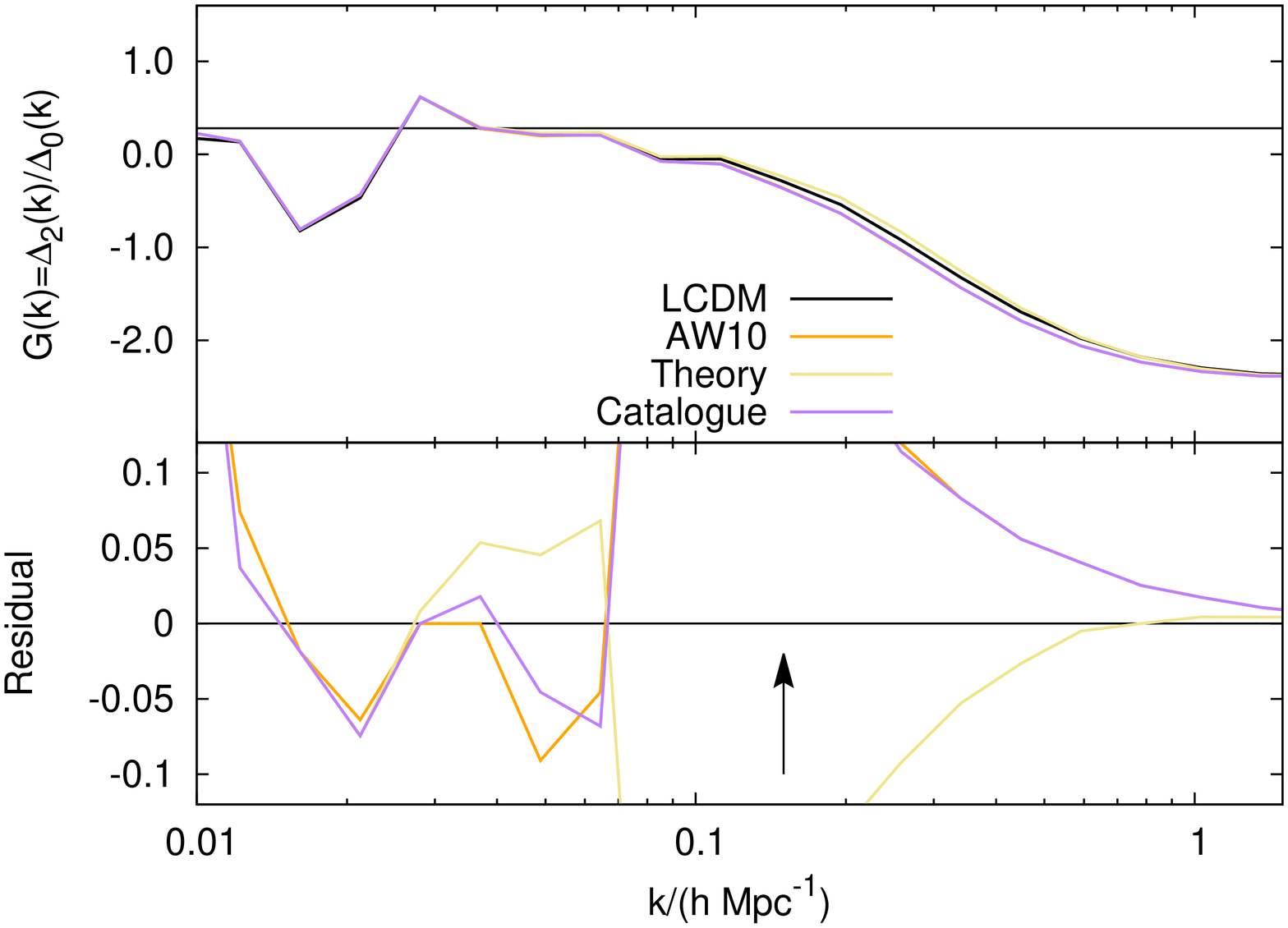}
\caption{The value of $G(k)=\Delta^2_2(k)/\Delta^2_0(k)$ recovered from the simulations before and after scaling. In each case the non-linear scale is shown with an arrow. The top left panel shows the case of the full particle distribution where methods of regurgitation and restructuring have been used to alter halo interiors. Here the residual difference is small across the full range of scales, and is at the $5\%$ level on linear scales, with the non-linear tail being reproduced \new{at the $5\%$ level by all methods, in the region where the quadrupole is not passing through zero. The full range of scales is best matched by the restructuring method}. The top right panel shows the case of haloes themselves, whereas the bottom panel shows the case of reconstituted halo particles with both theoretical and catalogued velocity dispersions used in the reconstitution, in both cases a comparison is made with the original AW10 method, which performs well here because the residual BAO is effectively divided out and with it the error due to a slightly incorrect $b(M)$ relation. For haloes the value of $G(k)$ is reproduced at the $10\%$ level across all scales and when the halo particles are considered the match is at the $5\%$ level when the quadrupole is not passing through zero and the non-linear tail is fairly well recovered in this case.}
\label{fig:growth_rate_recovery}
\end{figure*}

A realistic goal of forthcoming galaxy redshift surveys will be to
measure the growth rate of cosmic structure with $\sim 1\%$ accuracy
(\eg \emph{DESI}, \emph{Euclid}). Can such a precision be attained
via simulations
that have undergone rescaling?
Fig. \ref{fig:growth_rate_recovery} shows the recovered
$G(k)=\Delta^2_2(k)/\Delta^2_0(k)$ ratio for: the full
rescaled particle distribution; the rescaled halo distribution; the
distribution of particles in haloes reconstituted from halo
catalogues.

At large scales for the case of the full matter distribution all approaches
are accurate to around $5\%$ out to $k=0.1\iMpc$. If haloes are restructured
this accuracy is extended out to $k=0.2\iMpc$. For the case of both the haloes
and halo particles large deviations are seen at the largest scales which remedy
themselves to the $7\%$ level for haloes, and $5\%$ level for halo particles, out to the non-linear scale.

In each plot the linear theory $G_\mathrm{lin}$ is shown
(equation \ref{eq:G_definition}); for the full matter distribution $G_\mathrm{lin}\simeq 0.55$.
and for the haloes, we expect
$G_\mathrm{lin}\simeq 0.37$ or $0.28$ for respectively the equal-weighted and mass-weighted cases
(taking into account the  effective bias
discussed in Section \ref{sec:method}).
Fig. \ref{fig:growth_rate_recovery} indicates that the rescaled simulations
are consistent with these growth rates but that large deviations are seen around the box scale,
although in the case of haloes a simulation
containing larger scales (thus more linear modes) would help to
support this conclusion.

At smaller scales, the results are most satisfactory in the case of
the halo distribution, where the error in the 
quadrupole-to-monopole ratio is relatively flat, at
typically around 4\%. There are occasional spikes to higher
values, but these are unlikely to disrupt the recovery of
the growth rate at this typical error rate. Conversely,
the results for the overall mass are less impressive, with errors exceeding 5\% for
a broad band of wavenumbers beyond the nonlinear scale.
Fortunately, this case is not of practical interest, whereas
we are most concerned with particles in haloes as a simple
proxy for a galaxy catalogue. Here, the last panel in
Fig. \ref{fig:growth_rate_recovery} shows that
usefully precise results to wavelengths large enough to
extract most of the cosmological information ($k\simeq 0.3\iMpc$)
would require improved treatment of the internal halo
velocity dispersion, with the rescaled values and pure
theoretical values bracketing the truth.

In a practical survey analysis one could
marginalize over velocity dispersion nuisance parameters, but 
this would diminish the statistical power of the high-$k$ data.
Some encouragement comes from the fact that the
non-linear portion of $G(k)$ for haloes is particularly well matched,
which is related to the fact that they lack a strongly non-linear
FOG. It is plausible that future surveys may attempt to target only halo central
galaxies in order to mitigate the effect of FOG in this way, or to use other
weighting schemes (\eg \citealt{Seljak2009}).

In any case, the example illustrated here undoubtedly makes
greater demands on the method than would be encountered in most
practical applications, where the range of parameter space to
be spanned would consist of models that are closer to
$\Lambda$CDM. The above errors would undoubtedly decline in
proportion to the distance between models in parameter space.
This in itself could create new problems: if the degree of
rescaling is very small, then there will be little chance of
finding an output at an input redshift $z$ suitable to match
a desired target $z'$. Instead, we may have to accept that
the original simulation defines a grid in $z'$ for the target
cosmology, over which we have no choice. But provided the
outputs are reasonably well spaced, this need not be a problem:
construction of a light-cone
mock galaxy sample from HOD galaxies populating the snapshots would
proceed in much the same way, independent of the exact set of $z'$ values.

\section{Summary and Conclusions}
\label{sec:summary}

In this paper we have considered the rescaling of simulated
dark-matter haloes in order to create catalogues of haloes that are
characteristic of an altered cosmological model, focusing on the
behavior of the method in redshift space. This extends our previous
work in real space (Mead \& Peacock 2014; MP14), which was based on
the original proposal of Angulo \& White (2010; AW10). MP14
demonstrated that rescaling of simulations could be made to work in
real space directly from halo catalogues, without needing to
manipulate the full particle data. In addition to the resulting advantage
in speed and storage, this approach also allows improvements over
AW10 in terms of the dependence of bias on halo mass and the
small-scale clustering, which is affected by internal changes to
halo structure.

For these reasons, it is clearly of interest to see if the
MP14 approach also works acceptably in redshift space, allowing it
to be used as the basis for a complete approach to the construction
of simulated surveys. We have therefore carried out the extreme test of
rescaling a halo catalogue generated
from a matter-only \tcdm simulation into that of a more standard \lcdm
model. 
In redshift space the MP14 method works well at the level of
the monopole, as in the original AW10 case. For the full particle
distribution the redshift-space monopole was recovered at the $2\%$
level up to $k=0.1\iMpc$ and to $3\%$ to smaller scales if halo
internal properties are also manipulated. Rescaling also worked well
when applied to haloes in redshift space, although the improvements
gained from using a biased displacement field are less marked because
redshift space mixes in the unbiased velocity field. The monopole
agrees at the few $\%$ level out to $k=1\iMpc$, although the numerical results
show some
deviations at the very largest scales investigated and the origin of these is not known.
For reconstituted haloes, the monopole
power spectrum was recovered at the $1\%$ level up to $k=0.1\iMpc$ if
a biased displacement field is used and the agreement is at the $3\%$
level up to $k=1\iMpc$ if catalogued dispersions are also rescaled.

The quadrupole to monopole ratio, $G(k)$, is of especial
interest because it can be used in the linear
regime to infer the growth rate of cosmic
structure. We found that this could be recovered at the
typical level of around 5\% for haloes themselves, even
out to wavenumbers as high as $k=1\iMpc$. This is most
encouraging given how far our input $\tau$CDM model is from anything
close to $\Lambda$CDM, and it suggests that the required
1\% precision should readily be obtained for models that
are allowed by existing cosmological constraints.

But for particles in haloes (an example of a galaxy HOD, albeit
an unrealistic one), the high-$k$ precision in $G(k)$ is
considerably poorer, reflecting the difficulty in achieving
an appropriate level for the internal halo velocity dispersions
using simple scaling recipes. Again this problem will diminish
(at least in the case of rescaling `observed' dispersions) as the
input model approaches closer to the target.
There are several routes by which the modelling of the
halo-particle power spectrum may be improved -- either via
more sophisticated dynamical modelling of the halo
velocity dispersion, or via weighting
schemes that suppress this component of the FOG.
We therefore see the current results as giving encouragement
that halo rescaling will be a practically useful modelling method in
the era when data on redshift-space distortions reaches the 1\% level.

\section*{Acknowledgements}

AJM acknowledges the support of an STFC studentship and support from the European Research Council under the EC FP7 grant number 240185.

\label{lastpage}
\setlength{\bibhang}{2.0em}
\setlength\labelwidth{0.0em}
\bibliographystyle{mn2e}
\bibliography{mead}
\end{document}